\documentclass[amsmath,amssymb,aps,prc,twocolumn,superscriptaddress,showpacs,preprintnumbers]{revtex4-1}
\usepackage[dvipdfmx]{graphicx}
\usepackage{multirow}
\usepackage{times}
\usepackage[dvipdfmx]{color}
\allowdisplaybreaks[1]
\usepackage[normalem]{ulem}  % \sout{old text} for strikeout
\renewcommand\sout{\bgroup \color{red} \ULdepth=-.5ex \ULset}
\begin{document}
\preprint{KUNS-2569}

\title{
  Derivation of second-order relativistic hydrodynamics for reactive multi-component systems
}

\author{Yuta Kikuchi}
\email[]{kikuchi@ruby.scphys.kyoto-u.ac.jp}
%\thanks{}
\affiliation{Department of Physics, Faculty of Science, Kyoto University,
Kyoto 606-8502, Japan.}
%%%%%
\author{Kyosuke Tsumura}
\email[]{kyosuke.tsumura@fujifilm.com}
\affiliation{Analysis Technology Center,
  Research \& Development Management Headquarters,
  Fujifilm Corporation,
  Kanagawa 250-0193, Japan.}
%\thanks{}
%%%%%
\author{Teiji Kunihiro}
\email[]{kunihiro@ruby.scphys.kyoto-u.ac.jp}
\affiliation{Department of Physics, Faculty of Science, Kyoto University,
Kyoto 606-8502, Japan.}
%\thanks{}

\date{\today}

%%%%%%%%%%%%%%%%%%%%%%%%%%%%%%%%%%%%%%%%%%%%%%%%%%%%%%%%%%%%%%%%%%%%%
\begin{abstract}
We derive the second-order hydrodynamic equation for
reactive multi-component systems 
from the relativistic Boltzmann equation.
In the reactive system, 
particles can change their species under the restriction of the imposed conservation laws
during the collision process.
Our derivation is based on the renormalization group (RG) method, in which
the Boltzmann equation is solved in an organized perturbation method
as faithfully as possible and possible secular terms are resummed away by 
a suitable setting of the initial value of the distribution function.
The microscopic formulae of the relaxation times and the lengths are
explicitly given as well as those of the transport coefficients for the reactive multi-component
system.
The resultant hydrodynamic equation with these formulae has nice properties that
it satisfies the positivity of the entropy production 
rate and the Onsager's reciprocal theorem, which ensure the validity of our derivation.
\end{abstract}
%%%%%%%%%%%%%%%%%%%%%%%%%%%%%%%%%%%%%%%%%%%%%%%%%%%%%%%%%%%%%%%%%%%%%

\maketitle

%%%%%%%%%%%%%%%%%%%%%%%%%%%%%%%%%%%%%%%%%%%%%%%%%%%%%%%%%%%%%%%%%%%%%
%\setcounter{equation}{0}
\section{
  Introduction
}
\label{sec:sec1}
Under the highly extreme condition realized in the ultra-relativistic heavy ion collision experiments 
at the Relativistic Heavy Ion Collider (RHIC) at the Brookhaven National Laboratory (BNL) 
and the Large Hadron Collider (LHC) at CERN, the quark-gluon plasma (QGP)  
has been most probably created \cite{Shuryak:2003xe,Gyulassy:2004zy}. 
To reveal the properties of QGP created there, 
the hydrodynamic model is utilized to study the time development of the extremely hot and dense matter. 
In particular, the large hadronic elliptic flow ($v_2$) observed 
in the experiments indicates that QGP is described by the relativistic hydrodynamic equation 
with a tiny viscosity, which may imply the creation of the strongly coupled matter \cite{Bass:2000ib,Teaney:2000cw,Teaney:2003kp,Hirano:2005wx,Hirano:2005xf,Nonaka:2006yn,Baier:2006um,
Baier:2006sr,Baier:2006gy,Romatschke:2007jx,Romatschke:2007mq,Bozek:2011gq,Hirano:2012kj};
however, see also Refs.~\cite{Asakawa:2006tc,Asakawa:2006jn}.
Recent studies take into account such small viscous effects on the hydrodynamic expansion of the hot and dense matter to reveal the properties of QGP quantitatively, where a relativistic dissipative hydrodynamic equation is an indispensable tool.
A naive relativistic extension of the Navie-Stokes equation, however, has fundamental problems 
such as ambiguity of flow velocity, existence of unphysical instabilities, 
and lack of causality, and we need to introduce the second-order hydrodynamic equation 
to avoid such problems. 
Moreover, in the expanding object created in the experiments there are high density core which shows a nearly perfect fluidity and dilute corona in the peripheral region. In the latter part, more microscopic dynamics should be incorporated to the hydrodynamics since the viscous effect is too large to apply the naive viscous hydrodynamics \cite{Hirano:2005wx}. The second-order hydrodynamics actually include the mesoscopic dynamics and useful to analyze the hydrodynamic behavior of the spatially inhomogeneous matter \cite{Israel:1976tn,Israel:1979wp,Huovinen:2008te,Molnar:2009pq,El:2009vj,Pu:2009fj,Bouras:2010hm,Denicol:2010xn,Denicol:2010br,Tsumura:2012gq,Tsumura:2012kp,Denicol:2012cn,Jaiswal:2013npa,
Tsumura:2013uma,Tsumura:2015fxa}.
%%%%%

The way of formulation of the second-order hydrodynamics is, 
however, controversial and many kinds of equation are proposed. 
For instance, the Israel-Stewart equation, which is a kind of the second-order hydrodynamic equation 
widely used for the analyses of time development of the relativistic heavy ion collision or QGP, 
does not have established validity because ambiguous assumptions are imposed to derive it. 
In fact, it has been shown that the solutions are different from those 
of the relativistic Boltzmann equation quantitatively \cite{Huovinen:2008te,Molnar:2009pq,El:2009vj}.

Recently, to eliminate the ambiguity in the derivation of hydrodynamics and perform systematic formulation, 
the renormalization group (RG) method
 \cite{Chen:1994zza,Chen:1995ena,Kunihiro:1995zt,Kunihiro:1996rs,kunihiro1998dynamical,
Kunihiro:1997uy,Kunihiro:1998jp,Boyanovsky:1998aa,Ei:1999pk,Boyanovsky:1999cy,Hatta:2001ui,Boyanovsky:2003ui,Kunihiro:2005dd}
 has been applied to derive the second-order hydrodynamic equation
for the single-component system
both in the non-relativistic and relativistic cases
\cite{Tsumura:2012gq,Tsumura:2012kp,Tsumura:2013uma,Tsumura:2015fxa} 
as an extension of the first-order case \cite{Tsumura:2007ji}:
In this method, the Boltzmann equation is faithfully solved 
to obtain the distribution function around the local equilibrium state without imposing any ansatz
in a perturbation theory and the secular terms are resummed away 
into the would-be integral constants, which constitute the slow variables, 
i.e., the hydrodynamical variables.
It is also understood that, in the context of the reduction theory of dynamical systems 
\cite{guckenheimer1983nonlinear,kuramoto1989reduction}, 
the hydrodynamical variables constitute the natural coordinates of  
 the invariant/attractive manifold in the 
functional space spanned by the distribution function.
Then the renormalization group equation\cite{Chen:1994zza,Chen:1995ena}
 or envelope equation\cite{Kunihiro:1995zt,Kunihiro:1996rs} gives
the evolution equation of the hydrodynamical variables, i.e., the hydrodynamic equation,
 after an averaging of the distribution function
\cite{Hatta:2001ui,Kunihiro:2005dd,Tsumura:2007ji,Tsumura:2012gq,Tsumura:2013uma,Tsumura:2015fxa}.
It has been shown that the resultant second-order relativistic hydrodynamic equation is 
causal and stable \cite{Tsumura:2012gq,Tsumura:2012kp,Tsumura:2013uma,Tsumura:2015fxa},
the microscopic expressions for the transport coefficients 
coincide with those obtained by Chapman-Enskog method, and 
those of the relaxation times are different from any other previous ones
but allow physically natural interpretations.
Their numerical values are indeed different
from those by the moment method \cite{Tsumura:2015fxa}.

In reality, the QGP is not a single-component system but 
reactive multi-component system composed of  quarks, antiquarks and gluons.
Then  it is  imperative to  derive the multi-component relativistic 
hydrodynamic equation for a realistic description of the QGP.
In this paper, we apply the RG method to the derivation of the second-order 
relativistic hydrodynamic equation for the reactive 
multi-component systems as an extension of Ref.~\cite{Tsumura:2015fxa} for the single-component system.
In the reactive system, the species of the particles can be changed through the 
collision process in a way that the imposed conservation laws are satisfied:
We restrict the collision to be a  2 to 2 process for simplicity.
We prove that the positive definiteness of the entropy production rate and the Onsager's reciprocal
relation are readily satisfied by using our microscopic expressions of the transport coefficients.
Furthermore, we show the causality and stability of the resultant equation in Ref.~\cite{Tsumura:2015fxa}.
All these confirmation strongly 
suggest our formulation based on the RG method.

Historically, Prakash et al. \cite{Prakash:1993bt} first derived 
the second-order hydrodynamics for relativistic gas mixture as an extension of 
the Israel-Stewart equation and investigated the properties of hot hadronic matter.
Then, there have been some attempts to derive the hydrodynamics for a mixture \cite{Monnai:2010qp,El:2010mt,El:2012ka}.
Among them,
the work by Monnai and Hirano \cite{Monnai:2010qp} is 
of great importance in our point of view:
They improved  the Israel-Stewart's fourteen moment method 
by additionaly imposing 
two physically natural conditions, i.e., the positive-definiteness of the entropy production rate 
and the Onsager's reciprocal relation  for the transport coefficients 
to determine the functional form for the distribution function together with 
a careful order counting with respect to the Knudsen number.
Thus they arrived at
a multi-component relativistic hydrodynamic equation,
which happens to have the same form as that of ours to be given in the present paper. 
It may imply that 
the form of the multi-component relativistic hydrodynamic equation has been uniquely determined
provided that the equation should satisfy the positivity of the entropy production rate
and the reciprocal relations, irrespective of the microscopic forms of the transport coefficients.

This paper is organized as follows:
In Sec.~\ref{sec:sec2},
we briefly review the basics of the reactive multi-component Boltzmann equation.
In Sec.~\ref{sec:sec3},
we summarize theoretical foundations of the RG method.
In Sec.~\ref{sec:sec4},
we derive the second-order hydrodynamic equation for multi-component systems
by the RG method.
In Sec.~\ref{sec:sec5},
we discuss the basic properties of the resultant equation 
and the microscopic expressions of the relaxation times as well as 
the transport coefficients; a focus is put on
the positivity of the entropy production rate 
and the reciprocal properties of the mutual transport coefficients.
Section~\ref{sec:sec6} is devoted 
to a summary and concluding remarks.
In Appendix~\ref{sec:app1},
we work out to derive the functional forms of the excited modes
utilizing the faithful solution of the Boltzmann equation.
In Appendix~\ref{sec:app2},
as an example, we apply the RG method to solve the Van del Pol equation .
In Appendix~\ref{sec:app3},
the derivation of the relaxation equation is performed explicitly and the microscopic expressions of all the transport coefficients are given.

In this paper,
we use the natural unit, i.e., $\hbar = c = k_{\mathrm{B}} = 1$,
and the Minkowski metric $g^{\mu\nu} = \mathrm{diag}(+1,-1,-1,-1)$.

%%%%%%%%%%%%%%%%%%%%%%%%%%%%%%%%%%%%%%%%%%%%%%%%%%%%%%%%%%%%%%%%%%%%%

%%%%%%%%%%%%%%%%%%%%%%%%%%%%%%%%%%%%%%%%%%%%%%%%%%%%%%%%%%%%%%%%%%%%%
%\setcounter{equation}{0}
\section{
  Multi-component Boltzmann equation
}
\label{sec:sec2}

In this section,
we make a brief review of the relativistic transport theory based on the Boltzmann equation for reactive multi-component systems \cite{de1980relatlvlatlc}.
We consider a fluid composed of $N$ species with $M$ conserved currents. Let $q_k^A$ denote the charge of the $k$-th species associated with the $A$-th conserved current, where $k$ runs from $1$ to $N$ and $A$ from $1$ to $M$. For instance, the electromagnetic U(1) charge of an electron is written as $q_\mathrm{e}^{\mathrm{U(1)}}=-1$. 
The conserved currents and the energy-momentum tensors are written as
\begin{align}
 N_A^{\mu}&=\sum_{k=1}^N q_k^A\int\mathrm{d}p_k\,p_k^{\mu}f_{k,p_k}(x),
 \label{eq:particle_current}
 \\
 T^{\mu\nu}&=\sum_{k=1}^N \int\mathrm{d}p_k\,p_k^{\mu}p_k^{\nu}f_{k,p_k}(x),
 \label{eq:energy-momentum_tensor}
\end{align}
where $f_{k,p_k}(x)$ denotes the distribution function of the $k$-th species whose four-momentum is $p_k=(\sqrt{m_k^2+\boldsymbol{p}_k^2},\boldsymbol{p}_k)$
with $\boldsymbol{p}_k$ being the spatial components of the four momentum $p_k^\mu$.

Near-equilibrium behavior of the one-body distribution function may be described by the following multi-component relativistic Boltzmann equation \cite{de1980relatlvlatlc,cercignani2002relativistic}:
\begin{align}
 p_k^{\mu}\partial_{\mu}f_{k,p}(x)=\sum_{l=1}^NC_{kl}[f]_{p_k}(x),
 \label{MBE}
\end{align}
with the collision integral
\begin{align}
 &C_{kl}[f]_{p_k}(x)
 \nonumber\\
 &=\frac{1}{2}\sum_{i,j=1}^N\int\mathrm{d}p_l\mathrm{d}p_i\mathrm{d}p_j
 \nonumber\\
 &\times[f_{i,p_i}f_{j,p_j}(1+a_k f_{k,p_k})(1+a_l f_{l,p_l})\mathcal{W}_{ij | kl}
 \nonumber\\
 &-f_{k,p_k}f_{l,p_l}(1+a_i f_{i,p_i})(1+a_j f_{j,p_j})\mathcal{W}_{kl | ij}].
\end{align}
Here, 
we have suppressed the arguments $x$,
and abbreviated an integration measure as 
$\mathrm{d}p=\mathrm{d}^3\boldsymbol{p}/[(2\pi)^3 p^0]$.
$\mathcal{W}_{kl | ij}$ is
the transition probability due to the microscopic two-particle interaction
with the energy-momentum conservation
\begin{align}
  \label{eq:ChapA-2-1-004}
  \mathcal{W}_{ij | kl} \propto \delta^4(p_i + p_j - p_k - p_l)\prod_{A=1}^M\delta_{q^A_i + q^A_j , q^A_k + q^A_l},
\end{align} 
and the symmetry properties
\begin{align}
  \label{eq:sym1}
  0 &= \mathcal{W}_{ij | kl} - \mathcal{W}_{ji | lk},
  \\
  \label{eq:sym2}
  0 &= \sum_{i,j=1}^N\int\mathrm{d}p_i\mathrm{d}p_j \big[\mathcal{W}_{ij | kl} 
  -  \mathcal{W}_{kl | ij}\big]. 
\end{align}
The second relation is due to the unitarity of the scattering process \cite{de1980relatlvlatlc}, but note that $\mathcal{W}_{ij | kl} \ne \mathcal{W}_{kl | ij}$ owing to the existence of a threshold of each reaction. $a_k$ represents the quantum statistical effect, i.e., $a_k=+1$ for boson, 
$a_k=-1$ for fermion, and $a_k=0$ for the classical Boltzmann gas.
In the collision processes described by this collision integral, we consider only binary collisions which conserve the charge before and after collisions. We, however, allow reactions, i.e., the particle numbers of each component do not need to be conserved in a collision process. The allowance of the change of the total number of particles during each scattering process is of physical importance but it is left as a future project. When only elastic scattering are allowed, we further impose $q_k^A=\delta_{k,A}$ and 
$\mathcal{W}_{kl | ij}=(1-\delta_{kl}/2)(\delta_{ki}\delta_{lj}+\delta_{kj}\delta_{li})\mathcal{W}_{kl}$, where $\mathcal{W}_{kl}$ is the transition probability for the elastic scattering. 

\begin{figure}[t]
 \begin{center}
 \includegraphics[width=8cm]{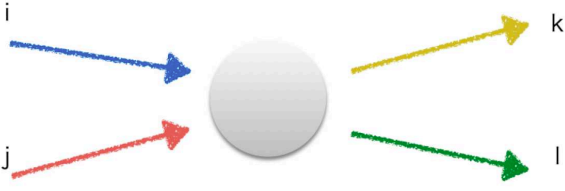}
 \caption{Schematic figure of a multi-component collision. Though only binary collision occur, kinds of particles may change during each scattering process with charges conserved, i.e., $q_i^A+q_j^A=q_k^A+q_l^A$.}
 \label{fig1}
 \end{center}
\end{figure}

On account of the symmetry properties \eqref{eq:sym1} and \eqref{eq:sym2},
the following identity for the collision integral for an arbitrary vector $\varphi_{k,p_k}$ \cite{footnote:vector} can be shown,
\begin{align}
  \label{eq:ChapA-2-1-006}
  &\sum_{k,l=1}^N\int\mathrm{d}p_k \varphi_{k,p_k} C_{kl}[f]_{p_k}
  \nonumber\\
  &= \frac{1}{4}\sum_{i,j,k,l=1}^N
  \int\mathrm{d}p_i\mathrm{d}p_j\mathrm{d}p_k\mathrm{d}p_l
  \nonumber\\
  &\times  (\varphi_{k,p_k}+\varphi_{l,p_l}- \varphi_{i,p_i} -\varphi_{j,p_j})
  \mathcal{W}_{ij | kl}
  \nonumber\\
  &\times f_{i,p_i}f_{j,p_j}(1+a_k f_{k,p_k})(1+a_k f_{l,p_l}).
\end{align}
Substituting $(q_k^A,p_k^\mu)$ into $\varphi_{k,p_k}$ in Eq.~(\ref{eq:ChapA-2-1-006}),
we find that $(q_k^A,p_k^\mu)$ are collision invariants satisfying
\begin{align}
  \label{eq:ChapA-2-1-007}
  \sum_{k,l=1}^N q_k^a\int\mathrm{d}p_k C_{kl}[f]_{p_k} &= 0,
  \\
  \label{eq:ChapA-2-1-008}
  \sum_{k,l=1}^N\int\mathrm{d}p_k p_k^\mu C_{kl}[f]_{p_k} &= 0,
\end{align}
which mean the charge and energy-momentum conservation in the collision process,
respectively.
These two equations leads to the balance equations
\begin{align}
  \label{eq:ChapA-2-1-009}
  \partial_\mu N_A^\mu
  &= 0,
  \\
  \label{eq:ChapA-2-1-010}
  \partial_\nu T^{\mu\nu}
  &= 0.
\end{align}
It should be noted, however, that
any dynamical properties are not contained in these equations
unless the evolution of $f_{k,p_k}$
has been obtained as a solution to Eq. \eqref{MBE}.

In the Boltzmann theory,
the entropy current may be defined \cite{de1980relatlvlatlc} by
\begin{align}
  \label{eq:ChapA-2-1-011}
  s^\mu 
  &\equiv - \sum_{k=1}^N\int\mathrm{d}p_k \, p_k^\mu 
  \Bigg[ f_{k,p_k}\ln  f_{k,p_k} 
  \nonumber\\
  &- \frac{(1+a_kf_{k,p_k})\ln(1+a_kf_{k,p_k})}{a_k} \Bigg].
\end{align}
The entropy-production rate reads
\begin{align}
  \partial_\mu s^\mu 
  &=  - \sum_{k=1}^N\int\mathrm{d}p_k\, p_k^\mu \partial_\mu f_{k,p_k} \ln \left(\frac{f_{k,p_k}}{1+a_k f_{k,p_k}}\right)
  \nonumber\\
  &=  - \sum_{k,l=1}^N\int\mathrm{d}p_k\, C_{kl}[f]_{p_k} \ln \left(\frac{f_{k,p_k}}{1+a_k f_{k,p_k}}\right),
  \label{entropy_prod1}
\end{align}
because of Eq.~\eqref{MBE}.
One sees that the entropy production rate vanishes only if $\ln (f_{k,p_k}/(1+a_kf_{k,p_k}))$ is a collision invariant,
i.e., 
$\ln (f_p/(1+af_p)) = \varphi_{k,p_k} = \sum_{A=1}^M q_k^A\alpha_A(x) + p^\mu \beta_\mu(x)$.
One thus finds \cite{de1980relatlvlatlc,cercignani2002relativistic} that
the entropy-conserving distribution function can be parametrized as
\begin{align}
 f_{k,p_k}
 &=\left[\exp\left(\frac{p_k\cdot u}{T}-\sum_{A=1}^M q_k^A\frac{\mu_A}{T}\right)-a_k\right]^{-1}
 \nonumber\\
 &\equiv f_{k,p_k}^{\mathrm{eq}}.
\end{align}
which is identified with the local equilibrium distribution function by interpreting
$T$, $\mu_A$, and $u^\mu$ 
as the local temperature,
chemical potentials associated with the conserved charges,
and flow velocity, respectively,
with the normalization 
\begin{align}
  \label{eq:normalization_of_flow}
  u^\mu  u_\mu = 1.
\end{align}
We see that the collision integral identically vanishes 
for the local equilibrium distribution $f^{\mathrm{eq}}_{k,p_k}$ as
\begin{align}
  \label{eq:ChapA-2-1-015}
  C_{kl}[f^{\mathrm{eq}}]_{p_k} = 0,
\end{align}
owning to the detailed balance
\begin{align}
 &\sum_{i,j=1}^N\int\mathrm{d}p_i\mathrm{d}p_j
 [f_{i,p_i}f_{j,p_j}(1+a_k f_{k,p_k})(1+a_l f_{l,p_l})\mathcal{W}_{ij | kl}
 \nonumber\\
 &-f_{k,p_k}f_{l,p_l}(1+a_i f_{i,p_i})(1+a_j f_{j,p_j})\mathcal{W}_{kl | ij}]
 \nonumber\\
 &=\sum_{i,j=1}^N\int\mathrm{d}p_i\mathrm{d}p_j \mathcal{W}_{ij | kl}
 [f_{i,p_i}f_{j,p_j}(1+a_k f_{k,p_k})(1+a_l f_{l,p_l})
 \nonumber\\
 &-f_{k,p_k}f_{l,p_l}(1+a_i f_{i,p_i})(1+a_j f_{j,p_j})]
 \nonumber\\
 &=0,
\end{align}
where the symmetry property \eqref{eq:sym2} and the conservations of energy-momentum and charges (\ref{eq:ChapA-2-1-004}) have been used in the first and second equality, respectively.

%%%%%%%%%%%%%%%%%%%%%%%%%%%%%%%%%%%%%%%%%%%%%%%%%%%%%%%%%%%%%%%%%%%%%

%%%%%%%%%%%%%%%%%%%%%%%%%%%%%%%%%%%%%%%%%%%%%%%%%%%%%%%%%%%%%%%%%%%%%
%\setcounter{equation}{0}
\section{
  Foundation of RG method
}
\label{sec:sec3}

In this section, we explain the renormalization group method used to identify slow variables of the original dynamics and obtain the reduced dynamics in terms of the slow variables \cite{Chen:1994zza,Kunihiro:1995zt}. Here, we explain the method following the discussion in \cite{Kunihiro:1996rs,Ei:1999pk}.
Let us consider the following $n$-dimensional differential equation:
\begin{align}
 \frac{\mathrm{d}\boldsymbol{X}}{\mathrm{d}t}=\boldsymbol{F}(\boldsymbol{X},t).
 \label{DE1}
\end{align}
We denote $\boldsymbol{W}(t)$ for the exact solution of Eq.~\eqref{DE1} though it is unspecified yet. By setting the exact solution $\boldsymbol{W}(t)$ as the initial condition, we solve Eq.~\eqref{DE1} perturbatively to obtain $\tilde{\boldsymbol{X}}(t;t_0,\boldsymbol{W}(t_0))$:
\begin{align}
 \label{eq:XBE}
 \frac{\mathrm{d}\tilde{\boldsymbol{X}}(t;t_0,\boldsymbol{W}(t_0))}{\mathrm{d} t}=\boldsymbol{F}(\tilde{\boldsymbol{X}}(t;t_0,\boldsymbol{W}(t_0)),t).
\end{align}
The basic concept of the renormalization group method is that the true solution of Eq.~\eqref{DE1} is independent of the initial time:
\begin{align}
 \tilde{\boldsymbol{X}}(t;t_0,\boldsymbol{W}(t_0))=\tilde{\boldsymbol{X}}(t;t'_0,\boldsymbol{W}(t'_0)),
\end{align}
which, by taking the limit $t'_0\to t_0$, leads to the RG equation
\begin{align}
 \left.\frac{\mathrm{d}\tilde{\boldsymbol{X}}(t;t_0,\boldsymbol{W}(t_0))}{\mathrm{d} t_0}\right|_{t_0=t}=0.
 \label{RG1}
\end{align}
Since the perturbative solutions $\tilde{\boldsymbol{X}}(t;t_0,\boldsymbol{X}(t_0))$ and $\tilde{\boldsymbol{X}}(t;t'_0,\boldsymbol{X}(t'_0))$ are only valid around $t\sim t_0$ and $t\sim t'_0$, respectively, the natural condition for $t$ is $t_0<t<t'_0$ which gives the condition $t_0=t$ after taking the limit  $t'_0\to t_0$.
Now, Eq.~\eqref{RG1} may be interpreted as a condition to construct the envelope of the family of curves which is represented by $\tilde{\boldsymbol{X}}(t;t'_0,\boldsymbol{X}(t'_0))$ with $t_0$ being a parameter characterizing curves \cite{Kunihiro:1995zt,Kunihiro:1996rs}.
By solving the RG equation, we can write down the RG improved solution, or envelope, as
\begin{align}
 \label{eq:XG}
 \boldsymbol{X}_{\mathrm{G}}(t)=\tilde{\boldsymbol{X}}(t;t_0,\boldsymbol{W}(t_0))|_{t_0=t}.
\end{align}

First, we check that the global solution \eqref{eq:XG} satisfies the original differential equation \eqref{DE1}.
Substituting Eq.~\eqref{eq:XG} into the left-hand side of Eq.~\eqref{DE1}, we have
\begin{align}
 &\frac{\mathrm{d}\tilde{\boldsymbol{X}}(t;t_0=t,\boldsymbol{W}(t))}{\mathrm{d}t}
 \nonumber\\
 &=\left.\frac{\mathrm{d}\tilde{\boldsymbol{X}}(t;t_0,\boldsymbol{W}(t_0))}{\mathrm{d}t}\right|_{t_0=t}
 +\left.\frac{\mathrm{d}\tilde{\boldsymbol{X}}(t;t_0,\boldsymbol{W}(t_0))}{\mathrm{d}t_0}\right|_{t_0=t}.
 \label{RG3}
\end{align}
Since $\tilde{\boldsymbol{X}}(t;t_0,\boldsymbol{W}(t_0))$ satisfies \eqref{eq:XBE} and because of the RG equation \eqref{RG1},
Eq.~\eqref{eq:XG} solves Eq.~\eqref{DE1}:
\begin{align}
 \frac{\mathrm{d}\boldsymbol{X}_G}{\mathrm{d}t}=\boldsymbol{F}(\boldsymbol{X}_G,t).
\end{align}

Next, we show the converse, namely, that $\tilde{\boldsymbol{X}}(t;t_0,\boldsymbol{W}(t_0))$ satisfies the RG equation \eqref{RG1}, provided that $X_G(t)$ given in Eq.~\eqref{eq:XG} solves the original differential equation \eqref{DE1}:
\begin{align}
 &\boldsymbol{F}(\boldsymbol{X}_G,t) 
 \nonumber\\
 &= \frac{\mathrm{d}\boldsymbol{X}_G}{\mathrm{d}t}
 \nonumber\\
 &=\left.\frac{\mathrm{d}\tilde{\boldsymbol{X}}(t;t_0,\boldsymbol{W}(t_0))}{\mathrm{d}t}\right|_{t_0=t}
 +\left.\frac{\mathrm{d}\tilde{\boldsymbol{X}}(t;t_0,\boldsymbol{W}(t_0))}{\mathrm{d}t_0}\right|_{t_0=t}.
\end{align}
Since $\tilde{\boldsymbol{X}}(t;t_0,\boldsymbol{W}(t_0))$ satisfies Eq.~\eqref{eq:XBE}, the satisfaction of the RG equation \eqref{RG1} is readily confirmed.

Therefore, the RG equation \eqref{RG1} and the original differential equation with the global solution inserted \eqref{eq:XBE} are equivalent. In the next section, we solve the Boltzmann equation following the latter method
although some elaboration is necessary in extending the invariant manifold so as to incorporating 
the excited modes \cite{Tsumura:2015fxa}.
In Appendix~\ref{sec:app2}, 
the equivalence of these two methods are demonstrated by 
explicitly working out an example with a limit cycle, although an incorporation 
of the excited modes is not made.

%%%%%%%%%%%%%%%%%%%%%%%%%%%%%%%%%%%%%%%%%%%%%%%%%%%%%%%%%%%%%%%%%%%%%

%%%%%%%%%%%%%%%%%%%%%%%%%%%%%%%%%%%%%%%%%%%%%%%%%%%%%%%%%%%%%%%%%%%%%
%\setcounter{equation}{0}
\section{
  Derivation of second-order hydrodynamics for multi-component systems
}
\label{sec:sec4}

The derivation of the hydrodynamic equation by the RG method explained in the last section consists of the following three steps:

\begin{enumerate}
 \item Solve the relativistic Boltzmann equation perturbatively with respect to small spatial inhomogeneity.
 \item Improve the perturbative solution to obtain the global solution with the renormalization group.
 \item Substitute the global solution into the Boltzmann equation and construct the moment equations, which give the hydrodynamic equation.
\end{enumerate}

In this section, we utilize the RG method to drive the second-order hydrodynamic equation for multi-component systems from the relativistic Boltzmann equation (\ref{eq:ChapA-3-1-015}). 
Then, we discuss some properties of the resultant equation,
focusing on the entropy production rate.
We shall see that the resultant equation readily satisfies the positivity of the entropy production rate and the Onager's reciprocal relation, which is the general properties of the non-equilibrium phenomena.

%---------------------------------------------------------------------------%
\subsection{
  Solving the Boltzmann equation to extract mesoscopic dynamics
}
\label{sec:ChapBr-4-2}

\subsubsection{
  An organized perturbation theory
}

To solve the Boltzmann equation perturbatively with respect to spatial inhomogeneity, which is assumed to be small since the system is in the hydrodynamic regime, we rewrite Eq.~\eqref{MBE} as
\begin{align}
  \label{eq:ChapA-3-1-015}
  \frac{\partial}{\partial \tau} f_{k,p_k}(\tau,\sigma)
  &=\frac{1}{p_k \cdot u} \sum_{l=1}^N C_{kl}[f]_{p_k}(\tau,\sigma)
  \nonumber\\
  &- \epsilon \frac{1}{p_k \cdot u}
  p_k^\mu\frac{\partial}{\partial\sigma^\mu} f_{k,p_k}(\tau,\sigma).
\end{align}
where $\epsilon$ is a bookkeeping parameter of the small spatial inhomogeneity corresponding to the Knudsen number = (mean free path) / (characteristic macroscopic length scale). 
After deriving the hydrodynamics, we shall put back $\epsilon =1$.
Here, we have introduced the temporal and spatial derivatives
\begin{align}
 \frac{\partial}{\partial \tau} = u^{\mu}\partial_{\mu},\ \ \  
 \frac{\partial}{\partial \sigma} = \nabla^{\mu} = (g^{\mu\nu}-u^{\mu}u^{\nu})\partial_{\nu},
\end{align}
respectively. $u^\mu$ is interpreted as the flow velocity satisfying $u_\mu u^\mu=1$, and $\Delta^{\mu\nu} \equiv g^{\mu\nu} - u^\mu u^\nu$. The present choice of the coordinate system leads to the relativistic hydrodynamics in the Landau-Lifshitz frame or the energy frame \cite{Tsumura:2012ss}.
From now on, we suppress the variable $\sigma$.
We expand the distribution function in the following perturbation series with respect to $\epsilon$:
\begin{align}
 \label{eq:pert_exp}
 &\tilde{f}(\tau;\tau_0)
 \nonumber\\
 &=\tilde{f}^{(0)}(\tau;\tau_0)+\epsilon \tilde{f}^{(1)}(\tau;\tau_0)+\epsilon^2 \tilde{f}^{(2)}(\tau;\tau_0)+\mathcal{O}(\epsilon^3).
\end{align}
Correspondingly, we expand the initial value as
\begin{align}
 &\tilde{f}(\tau=\tau_0;\tau_0)
 \nonumber\\
 &=f(\tau_0)
 \nonumber\\
 &=f^{(0)}(\tau_0)+\epsilon f^{(1)}(\tau_0)+\epsilon^2 f^{(2)}(\tau_0)+\mathcal{O}(\epsilon^3).
\end{align}
which is assumed to be an exact solution of Eq.~\eqref{eq:ChapA-3-1-015}.
Substituting them into Eq.~\eqref{eq:ChapA-3-1-015}, we obtain a series of equations, which we solve order by order. 

The zeroth order equation with respect to $\epsilon$ reads
\begin{align}
 \frac{\partial}{\partial\tau}\tilde{f}_{k,p_k}^{(0)}(\tau;\tau_0)
 =\frac{1}{p_k\cdot u}\sum_{l=1}^N C_{kl}[\tilde{f}^{(0)}]_{p_k}(\tau;\tau_0).
\end{align}

Since we are now considering a near-equilibrium system, the zeroth-order solution should satisfies
\begin{align}
 \frac{\partial}{\partial\tau}\tilde{f}_{k,p_k}^{(0)}(\tau;\tau_0)=0,
\end{align}
which leads to
\begin{align}
 \frac{1}{p_k\cdot u}\sum_{l=1}^N C_{kl}[\tilde{f}^{(0)}]_{p_k}(\tau;\tau_0)=0
\end{align}
Therefore, from Eq.~\eqref{eq:ChapA-2-1-015}
the local equilibrium distribution function is found to be the zeroth-order solution:
\begin{align}
 &\tilde{f}^{(0)}(\tau;\tau_0) = f^{\mathrm{eq}}(\tau_0).
\end{align}
Eq.~\eqref{eq:pert_exp} is understood as the expansion around the equilibrium distribution function $f^{\mathrm{eq}}$ and $\tilde{f}^{(i)}$ with $i\ge 1$ represents the non-equilibrium correction, which gives rise to dissipative effects. 

The first- and second-order equations read
\begin{align}
  \label{eq:1st-ordereq}
  \frac{\partial}{\partial\tau} \tilde{f}^{(1)}(\tau)
  &= f^{\mathrm{eq}}\bar{f}^{\mathrm{eq}} L 
  (f^{\mathrm{eq}} \bar{f}^{\mathrm{eq}})^{-1} 
  \tilde{f}^{(1)}(\tau) 
  + f^{\mathrm{eq}}\bar{f}^{\mathrm{eq}} F^{(0)},
  \\
  \label{eq:2nd-ordereq}
  \frac{\partial}{\partial\tau} \tilde{f}^{(2)}(\tau)
  &=f^{\mathrm{eq}} \bar{f}^{\mathrm{eq}} L  
  (f^{\mathrm{eq}} \bar{f}^{\mathrm{eq}})^{-1} \tilde{f}^{(2)}(\tau) 
  \nonumber\\
  &+ f^{\mathrm{eq}} \bar{f}^{\mathrm{eq}} K(\tau-\tau_0),
\end{align}
with the definitions of the following quantities
\begin{align}
  \label{eq:feq}
  &\bar{f}^{\mathrm{eq}}_{k,p_k} \equiv 1+a_k f^{\mathrm{eq}}_{k,p_k},
  \\
  \label{eq:F0}
  &F^{(i)}_{k,p_k}
  \equiv -(f^{\mathrm{eq}}_{k,p_k}\bar{f}^{\mathrm{eq}}_{k,p_k})^{-1}\frac{1}{p_k\cdot u}
  p_k\cdot\nabla\tilde{f}^{(i)}_{k,p_k},
  \\
   \label{eq:K}
  &K(\tau-\tau_0) \equiv
  F^{(1)}(\tau) + \frac{1}{2} B[\tilde{f}^{(1)},\tilde{f}^{(1)}](\tau),
  \\
  \label{eq:ChapBr-4-1-039}
  &B[\chi,\psi]_{k,p_k;m,q_m;n,r_n}
  \equiv -(f^{\mathrm{eq}}_{k,p_k}\bar{f}^{\mathrm{eq}}_{k,p_k})^{-1}
  \nonumber\\
  &\times\frac{\delta^2}{\delta f_{m,q_m}\delta f_{n,r_n}}
  \left.\left(\frac{1}{p_k\cdot u}\sum_{l=1}^NC_{kl}[f]_{k,p_k}\right)\right|_{f=f^{\mathrm{eq}}}
  \chi^{(1)}_{m,q_m} \psi^{(1)}_{n,r_n}.
\end{align}
and the linearized collision operator
\begin{align}
\label{eq:ChapBr-2-2-012}
  & L_{k,p_k;m,q_m}
  \equiv (f^{\mathrm{eq}}_{k,p_k}\bar{f}^{\mathrm{eq}}_{k,p_k})^{-1}
  \nonumber\\
  &\times\frac{\delta}{\delta f_{m,q_m}}
  \left.\left(\frac{1}{p_k\cdot u}\sum_{l=1}^N C_{kl}[f]_{k,p_k}\right)
  \right|_{f=f^{\mathrm{eq}}}
  f^{\mathrm{eq}}_{m,q_m}\bar{f}^{\mathrm{eq}}_{m,q_m}
  \nonumber\\
  &=-\sum_{l,i,j}
  \frac{1}{2p_k\cdot u}\int\mathrm{d}p_l\mathrm{d}p_i \mathrm{d}p_j
  \mathcal{W}_{ij | kl}
  \frac{\bar{f}^{\mathrm{eq}}_{l,p_l}f^{\mathrm{eq}}_{i,p_i}f^{\mathrm{eq}}_{j,p_j}}
  {f^{\mathrm{eq}}_{k,p_k}}
  \nonumber\\
  &\times\{\delta_{km}\delta(\bm{p_k}-\bm{q_m})+\delta_{lm}\delta(\bm{p_l}-\bm{q_m})
  \nonumber\\
  &-\delta_{im}\delta(\bm{p_i}-\bm{q_m})-\delta_{jm}\delta(\bm{p_j}-\bm{q_m})\},
\end{align}
With respect to the inner product defined as
\begin{align}
  \langle \psi , \chi \rangle
  \equiv \sum_{k=1}^N\int\mathrm{d}p_k (p_k\cdot u) 
  f^{\mathrm{eq}}_{k,p_k} \bar{f}^{\mathrm{eq}}_{k,p_k} \psi_{k,p_k} \chi_{k,p_k},
\end{align}
for arbitrary vectors $\psi_{k,p_k}$ and $\chi_{k,p_k}$,
the linearized collision operator $L_{k,p_k;m,q_m}$
is found to be self-adjoint and semi-negative definite:
\begin{align}
  \label{eq:self-adjoint}
  \langle \psi ,L \chi \rangle &= \langle L \psi , \chi {\rangle},
  \\
  \label{eq:non-positive}
  \langle \psi , L \psi\rangle &\le 0.
\end{align}
The operator $L$ has the $M+4$ eigenvectors belonging to the zero eigenvalue, i.e.,
\begin{align}
  \label{eq:ChapA-4-2-012}
  \varphi_{k,p_k}^\alpha \equiv \left\{
  \begin{array}{ll}
    \displaystyle{p_k^\mu,} & \displaystyle{\alpha = \mu=0,1,2,3,} \\[1mm]
    \displaystyle{q_k^A,}     & \displaystyle{\alpha = A+3=4,\cdots,M+3,}
  \end{array}
  \right.
\end{align}
which are the collision invariants
and satisfy
\begin{align}
  \label{eq:ChapA-4-2-013}
  \big[ L \varphi^\alpha \big]_{k,p_k} = 0.
\end{align}
We call $\varphi_{k,p_k}^\alpha$ the zero modes.
 
The $\mathrm{P_0}$-space is defined as the space spanned by the zero modes of the linearized collision operator, which govern the slow dynamics of the solution, and the $\mathrm{Q_0}$-space is defined as the complemental space of the $\mathrm{P_0}$-space in the solution space of the relativistic Boltzmann equation, which are spanned by excited modes.
The projection operator onto the $\mathrm{P_0}$-space and the $\mathrm{Q_0}$-space are given by
\begin{align}
  \label{eq:ChapA-4-2-014}
  \big[ P_0  \psi \big]_{k,p_k} 
  &\equiv
  \varphi_{k,p_k}^\alpha  \eta^{-1}_{\alpha\beta}
  \langle  \varphi^\beta ,\psi \rangle,
  \\
  \label{eq:ChapA-4-2-015}
  Q_0 &\equiv 1 - P_0,
\end{align}
where
$\eta^{-1}_{\alpha\beta}$ is the inverse matrix of
the the P${}_0$-space metric matrix $\eta^{\alpha\beta}$
defined by
\begin{align}
  \label{eq:ChapA-4-2-016}
  \eta^{\alpha\beta} \equiv \langle  \varphi^\alpha ,\varphi^\beta \rangle.
\end{align}

We further divide the Q${}_0$-space
into the P${}_1$- and Q${}_1$-space: P${}_1$-space is spanned by the {\it doublet modes} \cite{Tsumura:2013uma}, which consists of the sets of 
$(\hat{\Pi}_{k,p_k},\hat{J}^\mu_{A,k,p_k},\hat{\pi}^{\mu\nu}_{k,p_k})$ and $([L^{-1}\hat{\Pi}]_{k,p_k},[L^{-1}\hat{J}_A^\mu]_{k,p_k},[L^{-1}\hat{\pi}^{\mu\nu}]_{k,p_k})$
, while the Q${}_1$-space is the complement to the sum of the P${}_0$ and P${}_1$ spaces: Here,
$\hat{\Pi}$, $\hat{J}^\mu_A$, and $\hat{\pi}^{\mu\nu}$ are found to be the microscopic representations of the bulk viscosity, charge diffusion, and stress tensor, respectively, whose definitions are given in Appendix.~\ref{sec:app1}.
The correspondent projection operators are denoted by $P_1$ and $Q_1$, respectively.

Equations~\eqref{eq:1st-ordereq} and \eqref{eq:2nd-ordereq} can be solved much the same way as in \cite{Tsumura:2015fxa} using a vector representation: We refer to \cite{Tsumura:2015fxa} for the details. The first- and second-order solutions take the following forms:
\begin{align}
 \label{eq:first-order_solution}
  \tilde{f}^{(1)}(\tau)
  &=f^{\mathrm{eq}} \bar{f}^{\mathrm{eq}} 
  \Big[
  \mathrm{e}^{L(\tau-\tau_0)} \Psi + (\tau-\tau_0)P_0 F^{(0)}
  \nonumber\\
  &+ (\mathrm{e}^{L(\tau-\tau_0)} - 1)  L^{-1} Q_0  F^{(0)}
  \Big],
  \\
  \label{eq:2nd-ordersol}
  \tilde{f}^{(2)}(\tau)
  &=
  f^{\mathrm{eq}} \bar{f}^{\mathrm{eq}}
\Big[ (\tau-\tau_0)P_0
  + (\tau-\tau_0)\mathcal{G}(s)^{-1}P_1\mathcal{G}(s) Q_0
  \nonumber\\
  &- (1+(\tau-\tau_0)\partial/\partial s) Q_1 \mathcal{G}(s) Q_0\Big]\,K(s)\Big|_{s=0},
\end{align}
with the initial values
\begin{align}
 \label{eq:first-order_initialvalue}
  f^{(1)} &= f^{\mathrm{eq}} \bar{f}^{\mathrm{eq}} \Psi,
  \\
  \label{eq:2nd-orderinit}
  f^{(2)} &= -f^{\mathrm{eq}} \bar{f}^{\mathrm{eq}}
  Q_1\mathcal{G}(s)Q_0\,K(s)\Big|_{s=0},
\end{align}
After a lengthy calculation, which is worked out in App.~\ref{sec:app1}, we obtain the initial value $\Psi$ for the first-order equation:
\begin{align}
  \label{eq:Psi_in_RG}
  \Psi_{k,p_k} &=
  \Bigg[
    \frac{\big[ L^{-1}\hat{\Pi} \big]_{k,p_k}}{\langle \hat{\Pi},L^{-1}\hat{\Pi} {\rangle}}
  \Bigg] \Pi
  \nonumber\\
  &+\Bigg[
    \sum_{A,B=1}^M 3h\big[ L^{-1} \hat{J}_A^\mu \big]_{k,p_k}\langle \hat{J}^\nu, L^{-1}\hat{J}_{\nu} {\rangle}^{-1}_{AB}  
  \Bigg]  J_{B,\mu}
  \nonumber\\
  &+\Bigg[
    \frac{5\big[ L^{-1}\hat{\pi}^{\mu\nu} \big]_{k,p_k}}{\langle \hat{\pi}^{\rho\sigma}, L^{-1}\hat{\pi}_{\rho\sigma}{\rangle}}
  \Bigg] \pi_{\mu\nu},
\end{align}
where $h \equiv (e + P)/n$ is the enthalpy per particle
and the matrix notation is introduced as:
\begin{align}
 \langle \hat{J}^\mu,L^{-1}\hat{J}_\mu \rangle_{AB}
 \equiv \langle \hat{J}_A^\mu,L^{-1}\hat{J}_{B,\mu} \rangle,
\end{align}
and the $(A,B)$-component of its inverse matrix is denoted by 
$\langle \hat{J}^\mu,L^{-1}\hat{J}_\mu(s) \rangle^{-1}_{AB}$.
The following would-be $3M+6$ integral constants
have been introduced as the amplitudes of the nine vector fields: 
\begin{align}
  \Pi(\sigma ;\tau_0),\,\,\,J_A^\mu(\sigma ;\tau_0),\,\,\,\pi^{\mu\nu}(\sigma ;\tau_0).
\end{align}
We note that both $J_A^\mu$ and $\pi^{\mu\nu}$ are transverse and satisfy the equalities
\begin{align}
  \label{eq:constraint01}
  J_A^\mu &= \Delta^{\mu\nu}J_{A,\nu},
  \\
  \label{eq:constraint02}
  \pi^{\mu\nu} &= \Delta^{\mu\nu\rho\sigma}\pi_{\rho\sigma},
\end{align}
where $\Delta^{\mu\nu\rho\sigma}$ is a traceless symmetric projection operator 
\begin{align}
  \Delta^{\mu\nu\rho\sigma}
  \equiv 1/2(\Delta^{\mu\rho}\Delta^{\nu\sigma} + \Delta^{\mu\sigma}\Delta^{\nu\rho}
  - 2/3 \Delta^{\mu\nu}\Delta^{\rho\sigma}).
\end{align}
We note that
$\Pi$, $J_A^\mu$, and $\pi^{\mu\nu}$ will be interpreted
as the bulk pressure, thermal flux, and stress tensor, respectively.
We have also introduced a ``propagator" defined by 
\begin{align}
 \label{eq:propagator}
 \mathcal{G}(s)\equiv (L-\partial/\partial s)^{-1},
\end{align} 
which represents the temporal non-locality, i.e., non-Markovian effect.

Summing up
the perturbative solutions up to the second order with respect to $\epsilon$,
we have the full expression of
the initial value
and the perturbative solution around $\tau \sim \tau_0$ to the second order:
\begin{align}
  \label{eq:solution}
  \tilde{f}(\tau)
  &=
  f^{\mathrm{eq}}
+\epsilon f^{\mathrm{eq}} \bar{f}^{\mathrm{eq}}
\Big[(1+(\tau-\tau_0)L)  \Psi + (\tau-\tau_0) F_0 \Big] 
  \nonumber\\
  &+ \epsilon^2f^{\mathrm{eq}} \bar{f}^{\mathrm{eq}}
\Big[ (\tau-\tau_0)P_0+ (\tau-\tau_0)\mathcal{G}(s)^{-1}P_1\mathcal{G}(s)Q_0
  \nonumber\\
  &- (1+(\tau-\tau_0)\partial/\partial s)Q_1\mathcal{G}(s)Q_0\Big]\,K(s)\Big|_{s=0},
  \\
  \label{eq:initial}
  f &= f^{\mathrm{eq}}+\epsilon f^{\mathrm{eq}} \bar{f}^{\mathrm{eq}} \Psi
- \epsilon^2 f^{\mathrm{eq}} \bar{f}^{\mathrm{eq}} 
  Q_1 \mathcal{G}(s) Q_0 K(s)\Big|_{s=0},
\end{align}
which has completely the same structure as the one in the single-component case in the vector notation \cite{Tsumura:2015fxa}.

\subsubsection{
  RG improvement and moment equations
}
The perturbative solution (\ref{eq:solution}) possesses secular terms, which become divergent and clearly breaks down the perturbation as $|\tau-\tau_0|$ gets larger.
Improvement with the RG equation allows us to avoid such secular divergences and obtain the global solution \cite{Chen:1994zza,Chen:1995ena,Kunihiro:1995zt,Kunihiro:1996rs,Ei:1999pk,Hatta:2001ui,Kunihiro:2005dd}. 
By applying the RG equation to the local solution (\ref{eq:solution}),
we obtain the $\tau_0$-dependence of the would-be integral constants, which validate the approximate solution in a global domain:
\begin{align}
  \label{eq:RGeq}
  \frac{\mathrm{d}}{\mathrm{d}\tau_0}
  \tilde{f}_{k,p_k}(\tau,\sigma ;\tau_0) \Bigg|_{\tau_0 = \tau} = 0.
\end{align}
The RG equation (\ref{eq:RGeq}) indeed gives the equation of motion
governing the dynamics of the would-be $4M+10$ integral constants
$T(\sigma;\tau)$, $\mu_a(\sigma;\tau)$, $u^\mu(\sigma;\tau)$,
$\Pi(\sigma;\tau)$, $J_a^\mu(\sigma;\tau)$, and $\pi^{\mu\nu}(\sigma;\tau)$.
The global solution can be obtained as
the initial value (\ref{eq:initial}) 
\begin{align}
  \label{eq:envelope}
  f^{\mathrm{G}}(\tau) 
  &\equiv f(\tau_0 = \tau)\nonumber\\
  &= f^{\mathrm{eq}}(1 + \epsilon \bar{f}^{\mathrm{eq}}\Psi)\nonumber\\
  &- \epsilon^2f^{\mathrm{eq}} \bar{f}^{\mathrm{eq}}
  Q_1 \mathcal{G}(s)Q_0K(s)\Big|_{s=0}\Bigg|_{\tau_0 = \tau},
\end{align}
where the solution to Eq. (\ref{eq:RGeq}) is to be inserted.
We have derived the mesoscopic dynamics of the relativistic Boltzmann equation (\ref{eq:ChapA-3-1-015})
in the form of the pair of Eqs. (\ref{eq:RGeq}) and (\ref{eq:envelope}).
Let us remark on the global solution obtained here:
\begin{align}
  f^{\mathrm{G}} = f^{\mathrm{eq}}+\epsilon f^{\mathrm{eq}} \bar{f}^{\mathrm{eq}} \Psi + \epsilon^2 f^{(2)} + O(\epsilon^3),
\end{align}
with
\begin{align}
  \label{eq:2nd_global}
  f^{(2)} = -f^{\mathrm{eq}} \bar{f}^{\mathrm{eq}} 
  Q_1 \mathcal{G}(s) Q_0 K(s)\Big|_{s=0}.
\end{align}
Without loss of generality,
we can suppose that $\Psi$ contains no zero modes,
because such zero modes can be eliminated
by the redefinition of the zeroth-order initial value.
It means that
the possible existence of the zero modes in $\Psi$
would be renormalized into the local temperature, chemical potential, and flow velocity,
where the local frame is defined and fixed
by the flow velocity $u^{\mu}(\sigma; \tau_0)$.
This is a kind of the matching condition.
We note that the expression \eqref{eq:2nd_global} tells us that $f^{(2)}$ does not include P${}_1$- modes. Summarizing the above, we have the following orthogonality conditions
\begin{align}
 \label{eq:matching1}
 0&=\langle\varphi^\alpha, \Psi\rangle 
 =\langle\varphi^\alpha, (f^{\mathrm{eq}} \bar{f}^{\mathrm{eq}})^{-1}f^{(2)}\rangle,
 \\
 \label{eq:matching2}
 0&=\langle (\hat{\Pi},\hat{J}_A^\mu,\hat{\pi}^{\mu\nu}),(f^{\mathrm{eq}} \bar{f}^{\mathrm{eq}})^{-1}f^{(2)} \rangle
 \nonumber\\
 &=\langle L^{-1} (\hat{\Pi},\hat{J}_A^\mu,\hat{\pi}^{\mu\nu}),(f^{\mathrm{eq}} \bar{f}^{\mathrm{eq}})^{-1}f^{(2)} \rangle.
\end{align}

Although, in the usual RG method, the RG equation \eqref{eq:RGeq} describe the slow dynamics of the would-be integral constants, the insertion of the global solution \eqref{eq:envelope} into the Boltzmann equation \eqref{MBE} give the equivalent dynamics for the would-be integral constants.
Substituting Eq.~\eqref{eq:envelope} into Eq.~\eqref{MBE}, we have
\begin{align}
 &p_k^{\mu}\partial_{\mu}\big[f_{k,p_k}^{\mathrm{eq}}(1 + \epsilon \bar{f}_{k,p_k}^{\mathrm{eq}}\Psi_{k,p_k})\big]
 \nonumber\\
 &=\epsilon(p_k\cdot u)f_{k,p_k}^{\mathrm{eq}}\bar{f}_{k,p_k}^{\mathrm{eq}} L_{k,p_k;m,q_m}\Psi_{m,q_m}
  \nonumber\\
  &+\epsilon^2(p_k\cdot u)f_{k,p_k}^{\mathrm{eq}}\bar{f}_{k,p_k}^{\mathrm{eq}} L_{k,p_k;m,q_m}(f_{m,q_m}^{\mathrm{eq}} \bar{f}_{m,q_m}^{\mathrm{eq}})^{-1}f_{m,q_m}^{(2)}
  \nonumber\\
  &+\epsilon^2(p_k\cdot u) f_{k,p_k}^{\mathrm{eq}} \bar{f}_{k,p_k}^{\mathrm{eq}} \frac{1}{2} B_{k,p_k}[f^{(1)},f^{(1)}],
 \label{MBE_1st},
\end{align}
which is further reduced to simpler forms by taking 
the inner product with the zero modes $\varphi^\alpha_{k,p_k}$
and the excited modes $\big[ L^{-1} (\hat{\Pi},\hat{J}_A^\mu,\hat{\pi}^{\mu\nu}) \big]_{k,p_k}$
used in the definition of $\Psi_{k,p_k}$, respectively.
The averaging inner product with the zero modes leads to
\begin{align}
  \label{eq:ChapB-RHD1}
  &\sum_{k=1}^N \int\mathrm{d}p_k \varphi^{\alpha}_{k,p_k}
  \Bigg[(p_k\cdot u)\frac{\partial}{\partial\tau} + \epsilon p_k\cdot\nabla\Bigg]
  \nonumber\\
  &\times\Bigg[ f^\mathrm{eq}_{k,p_k} 
  ( 1 + \epsilon \bar{f}^{\mathrm{eq}}_{k,p_k} \Psi_{k,p_k}) \Bigg]\nonumber\\
  &= 0 + O(\epsilon^3),
\end{align}
and the equation reduced by the excited modes reads
\begin{align}
  \label{eq:ChapB-RHD2}
  &\sum_{k=1}^N\int\mathrm{d}p_k \big[ L^{-1}  (\hat{\Pi},\hat{J}_A^\mu,\hat{\pi}^{\mu\nu}) \big]_{k,p_k} 
  \Bigg[ (p_k\cdot u)  \frac{\partial}{\partial\tau}
  + \epsilon p_k\cdot\nabla \Bigg]
  \nonumber\\
  &\times\Bigg[ f^\mathrm{eq}_{k,p_k} (1 + \epsilon \bar{f}^{\mathrm{eq}}_{k,p_k} \Psi_{k,p_k}) \Bigg]
  \nonumber\\
  &= \epsilon \langle L^{-1} (\hat{\Pi},\hat{J}_A^\mu,\hat{\pi}^{\mu\nu}),
  L \Psi \rangle
  \nonumber\\
  &+ \epsilon^2 \frac{1}{2}\langle L^{-1} (\hat{\Pi},\hat{J}_A^\mu,\hat{\pi}^{\mu\nu}) ,
  B[\Psi,\Psi]\rangle
  + O(\epsilon^3).
\end{align}
Eqs. (\ref{eq:ChapB-RHD1}) and (\ref{eq:ChapB-RHD2}) give
the equations of motion governing $T$, $\mu_a$, $u^\mu$,
$\Pi$, $J_A^\mu$, and $\pi^{\mu\nu}$
in the P$_0$-space and P$_1$-space, respectively.
%---------------------------------------------------------------------------%

%---------------------------------------------------------------------------%
\subsection{
  Hydrodynamic equation
}
Now, we insert $\epsilon=1$ in Eqs.~\eqref{eq:ChapB-RHD1} and \eqref{eq:ChapB-RHD2}. Then,
Eq. \eqref{eq:ChapB-RHD1} is reduced to the following form
\begin{align}
  \label{eq:balanceeqbyRG}
  \partial_\mu J^{\mu\alpha}_{\mathrm{hydro}} = 0,
\end{align}
with
\begin{align}
  \label{eq:currentsbyRG}
  J^{\mu\alpha}_{\mathrm{hydro}}
  &\equiv \sum_{k=1}^N \int\mathrm{d}p_k p_k^\mu \varphi^\alpha_{k,p_k}
  f^\mathrm{eq}_{k,p_k}  ( 1 + \bar{f}^{\mathrm{eq}}_{k,p_k}\Psi_{k,p_k})
  \nonumber\\
  &= \left\{
  \begin{array}{ll}
    \displaystyle{e u^\mu u^\nu - (P + \Pi) \Delta^{\mu\nu} + \pi^{\mu\nu},} & \displaystyle{\alpha = \nu,} \\[2mm]
    \displaystyle{n_A u^\mu + J_A^\mu,} & \displaystyle{\alpha = A+3.}
  \end{array}
  \right.
\end{align}
Here, 
$n$, $e$, and $P$ denote the particle-number density, internal energy, and pressure, respectively,
whose microscopic representations
are given by
\begin{align}
 n_A &\equiv
 u_{\mu}N_A^{\mu}=\sum_{k=1}^N q_k^A\int\mathrm{d}p_k(p_k\cdot u)f_{k,p_k},
 \label{eq:Chap0-004}
 \\
 e &\equiv
 u_{\mu}u_{\nu}T^{\mu\nu}=\sum_{k=1}^N\int\mathrm{d}p_k(p_k\cdot u)^2f_{k,p_k},
 \label{eq:Chap0-005}
 \\
 P &\equiv
 -\frac{1}{3}\Delta_{\mu\nu}T^{\mu\nu}=-\frac{1}{3}\sum_{k=1}^N\int\mathrm{d}p\Delta_{\mu\nu}p_k^{\mu}p_k^{\nu}f^{\mathrm{eq}}_{k,p_k},
 \label{eq:Chap0-006}
\end{align}
We remark that
Eq. (\ref{eq:balanceeqbyRG})
is nothing but the balance equations
and $J^{\mu\nu}_{\mathrm{hydro}}$ and 
$J^{\mu\ A+3}_{\mathrm{hydro}}$ can be identified with
the energy-momentum tensor $T^{\mu\nu}$
and charge current $N_A^\mu$ given by Eqs. (\ref{eq:energy-momentum_tensor}) and (\ref{eq:particle_current}), respectively.

After a straightforward but lengthy manipulation worked out in Appendix.~\ref{sec:app3},
we can reduce Eq.~\eqref{eq:ChapB-RHD2} into the following relaxation equations:
\begin{widetext}
\begin{align}
  \label{eq:relax1}
  \Pi
  &= -\zeta\theta- \tau_\Pi \frac{\partial}{\partial\tau}\Pi - \sum_{a=1}^{M}\ell^a_{\Pi J}\nabla\cdot J_A
  \nonumber\\
  &+\kappa_{\Pi\Pi} \Pi \theta
  +\sum_{A=1}^{M}\kappa^{(1)A}_{\Pi J}J_{A,\rho}\nabla^\rho T
  +\sum_{A,B=1}^{M}\kappa^{(2)BA}_{\Pi J}J_{A,\rho}\nabla^\rho \frac{\mu_B}{T}
  +\kappa_{\Pi\pi}\pi_{\rho\sigma}\sigma^{\rho\sigma}
   \nonumber\\
  &+ b_{\Pi\Pi\Pi}\Pi^2 + \sum_{A,B=1}^{M}b_{\Pi JJ}^{AB}J_A^\rho J_{B,\rho} + b_{\Pi\pi\pi}\pi^{\rho\sigma}\pi_{\rho\sigma},
  \\
  \label{eq:relax2}
  J_A^\mu
  &= \sum_{B=1}^{M}\lambda_{AB}\frac{T^2}{h^2} \nabla^\mu \frac{\mu_{B}}{T}
  - \sum_{B=1}^{M}\tau_J^{AB} \Delta^{\mu\rho}\frac{\partial}{\partial\tau}J_{B,\rho}
  - \ell^A_{J\Pi}\nabla^\mu \Pi - \ell^A_{J\pi}\Delta^{\mu\rho} \nabla_\nu {\pi^\nu}_\rho
  \nonumber\\
  &+ \kappa^{(1)A}_{J\Pi}\Pi\nabla^\mu T 
  + \sum_{B=1}^{M}\kappa^{(2)AB}_{J\Pi}\Pi\nabla^\mu \frac{\mu_{B}}{T}
  + \sum_{B=1}^{M}\kappa^{(1)AB}_{JJ}J^\mu_B\theta
  + \sum_{B=1}^{M}\kappa^{(2)AB}_{JJ}J_{B,\rho}\sigma^{\mu\rho}+ \kappa^{(3)AB}_{JJ}J_{B,\rho}\omega^{\mu\rho}
  \nonumber\\
  &+ \kappa^{(1)A}_{J\pi}\pi^{\mu\rho}\nabla_\rho T
  + \sum_{B=1}^{M}\kappa^{(2)AB}_{J\pi}\pi^{\mu\rho}\nabla_\rho \frac{\mu_{B}}{T}
  \nonumber\\
  &+ \sum_{B=1}^{M}b^{AB}_{J\Pi J}\Pi J_B^\mu +  \sum_{B=1}^{M}b^{AB}_{JJ\pi}J_{B,\rho}\pi^{\rho\mu},
  \\
  \label{eq:relax3}
  \pi^{\mu\nu}
  &= 2\eta\sigma^{\mu\nu}
  - \tau_\pi \Delta^{\mu\nu\rho\sigma}\frac{\partial}{\partial\tau}\pi_{\rho\sigma}
  - \sum_{a=1}^{M}\ell^a_{\pi J}\nabla^{\langle\mu} J^{\nu\rangle}_a 
  \nonumber\\
  &+\kappa_{\pi\Pi}\Pi\sigma^{\mu\nu}
  + \sum_{A=1}^{M}\kappa^{(1)A}_{\pi J}J^{\langle\mu}_A\nabla^{\nu\rangle} T
  + \sum_{A,B=1}^{M}\kappa^{(2)BA}_{\pi J}J^{\langle\mu}_A\nabla^{\nu\rangle} \frac{\mu_B}{T}
  + \kappa^{(1)}_{\pi\pi}\pi^{\mu\nu}\theta
  + \kappa^{(2)}_{\pi\pi}\pi^{\lambda\langle\mu} {\sigma^{\nu\rangle}}_{\lambda}
  + \kappa^{(3)}_{\pi\pi}\pi^{\lambda\langle\mu} {\omega^{\nu\rangle}}_{\lambda}
  \nonumber\\
  &+ b_{\pi\Pi\pi} \Pi \pi^{\mu\nu}
  + \sum_{A,B=1}^{M}b^{AB}_{\pi JJ} J^{\langle\mu}_A J^{\nu\rangle}_B
  + b_{\pi\pi\pi} \pi^{\lambda\langle\mu} {\pi^{\nu\rangle}}_{\lambda}.
\end{align}
\end{widetext}
where $A^{\langle\mu\nu\rangle}\equiv\Delta^{\mu\nu\rho\sigma}A_{\rho\sigma}$ is a traceless symmetric tensor. The scalar expansion $\theta\equiv\nabla\cdot u$, the shear tensor $\sigma^{\mu\nu}\equiv\Delta^{\mu\nu\rho\sigma}\nabla_{\rho\sigma}$, the vorticity term $\omega^{\mu\nu} \equiv \frac{1}{2}  (\nabla^\mu u^\nu - \nabla^\mu u^\nu)$,
and many coefficients have been introduced, where explicit definitions are given in App.~\ref{sec:app3}. You also find the relaxation equation with $\epsilon$ explicitly shown in App.~\ref{sec:app3}.

Here, we write down the resultant microscopic representations of
the transport coefficients $\zeta$, $\lambda_{AB}$, and $\eta$, 
and relaxation times $\tau_\Pi$, $\tau_J^{AB}$, and $\tau_\pi$,
as follows:
\begin{align}
  \label{eq:TC1byRG}
  \zeta
  &\equiv -\frac{1}{T}\langle\hat{\Pi},L^{-1}\hat{\Pi}\rangle
  \nonumber\\
  &=\frac{1}{T}\int_0^{\infty}\mathrm{d}s\langle\hat{\Pi}(0),\hat{\Pi}(s)\rangle,
  \\
  \label{eq:TC2byRG}
 \lambda_{AB} 
 &\equiv \frac{1}{3T^2}\langle\hat{J}_A^{\mu},L^{-1}\hat{J}_{B,\mu}\rangle
 \nonumber\\
 &=-\frac{1}{3T^2}\int_0^{\infty}\mathrm{d}s\langle\hat{J}_A^{\mu}(0),\hat{J}_{B,\mu}(s)\rangle,
  \\
  \label{eq:TC3byRG}
  \eta 
  &\equiv -\frac{1}{10T}\langle\hat{\pi}^{\mu\nu},L^{-1}\hat{\pi}_{\mu\nu}\rangle
  \nonumber\\
  &=\frac{1}{10T}\int_0^{\infty}\mathrm{d}s
 \langle\hat{\pi}^{\mu\nu}(0),\hat{\pi}_{\mu\nu}(s)\rangle,
\end{align}
where we have defined the ``time-evolved'' vectors by
\begin{align}
  \label{eq:ChapA-5-2-008}
  &(\hat{\Pi}(s),\hat{J}^\mu_A(s),\hat{\pi}^{\mu\nu}(s))_{k,p_k}
  \nonumber\\
  &\equiv
  \sum_{m=1}^M\int\mathrm{d}q \big[ \mathrm{e}^{sL} \big]_{k,p_k;m,q_m}
  (\hat{\Pi},\hat{J}^\mu_A,\hat{\pi}^{\mu\nu})_{m,q_m}.
\end{align}
Here, we note the symmetry property: $\lambda_{AB}=\lambda_{BA}$.
The relaxation times have the expressions given by
\begin{align}
  \label{eq:RT1byRG}
  \tau_\Pi &\equiv - \frac{\langle \hat{\Pi}, L^{-2} \hat{\Pi} \rangle
  }{
  \langle \hat{\Pi}, L^{-1} \hat{\Pi} \rangle}
  =\frac{
    \int_0^\infty \mathrm{d}s\,s\langle \hat{\Pi}(0),\hat{\Pi}(s)
  \rangle
  }{
    \int_0^\infty \mathrm{d}s\langle \hat{\Pi}(0),\hat{\Pi}(s)
  \rangle
  },
  \\
  \label{eq:RT2byRG}
  \tau_{J}^{AB} &\equiv 
  -\sum_{C=1}^{M}\big<L^{-1}\hat{J}^\mu,L^{-1}\hat{J}_\mu\big>_{AC}
  \langle\hat{J}^\mu,L^{-1}\hat{J}_\mu\rangle^{-1}_{CB}
  \nonumber\\
  &=\sum_{C=1}^{M}
  \left(\int_0^\infty \mathrm{d}s\,s
  \langle \hat{J}^\mu(0),\hat{J}_\mu(s) \rangle \right)_{AC}
  \nonumber\\
  &\times \left(\int_0^\infty \mathrm{d}s
  \langle \hat{J}^\mu(0),\hat{J}_\mu(s) \rangle \right)^{-1}_{CB},
  \\
  \label{eq:RT3byRG}
  \tau_\pi &\equiv - \frac{\langle \hat{\pi}^{\mu\nu}, L^{-2} \hat{\pi}_{\mu\nu} \rangle
  }{
  \langle \hat{\pi}^{\rho\sigma} ,L^{-1} \hat{\pi}_{\rho\sigma} \rangle}
  =\frac{
    \int_0^\infty \mathrm{d}s\,s
    \langle \hat{\pi}^{\mu\nu}(0),\hat{\pi}_{\mu\nu}(s) \rangle
  }{
    \int_0^\infty \mathrm{d}s
    \langle \hat{\pi}^{\mu\nu}(0),\hat{\pi}_{\mu\nu}(s) \rangle
  },
\end{align}

We note that the cross correlations are generated by $\lambda_{AB}$, $\tau_J^{AB}$, and so on for $A\ne B$. The existence of the cross correlations is characteristic feature of the multi-component fluid. Owing to the terms including such transport coefficients, a diffusion of some charge influences that of the others. 

It is noteworthy that $\zeta$, $\lambda_{AA}$, $\eta$, $\tau_\Pi$, $\tau_J^{AA}$, and $\tau_\pi$ can be proved to be positive definite as a result of the semi-negative definiteness of the linearized collision operator $L$ as given in Eq.~\eqref{eq:non-positive}. 

Here, we prove the positive definiteness of the bulk viscosity $\zeta$ and the proof for the other coefficients are straightforward.
Since $L^{-1}$ is symmetric and negative definite when acting on the excited modes, there exists a real (lower triangular) matrix $U$, such that  $L^{-1}=-U^t U$, which is called Cholesky decomposition. By using the Cholesky decomposition, we have
\begin{align}
 \langle\hat{\Pi},L^{-1}\hat{\Pi}\rangle
 =-\langle\hat{\Pi},U^t U\hat{\Pi}\rangle
 =-\langle U\hat{\Pi},U\hat{\Pi}\rangle < 0
\end{align}
which show the positive definiteness of the bulk viscosity \eqref{eq:TC1byRG}.

Let us clarify the difference between our result and the previous attempts
 \cite{Prakash:1993bt,Monnai:2010qp,El:2010mt,El:2012ka} to derive the second-order 
hydrodynamic equation for relativistic multi-component systems. 
In \cite{Prakash:1993bt},
 the Israel-Stewart theory is simply extended to multi-component systems and hence 
the resultant equation is not free from the
drawbacks that the Israel-Stewart equation possessed, even apart
 from the shortcomings that the reactive effects are not included. In \cite{Monnai:2010qp}, 
though they also used the moment method, they obtained new second-order terms by performing
 the careful order counting with respect to the Knudsen number and indeed the resultant equation is consistent with ours. 
However, since the collision integral of the Boltzmann equation was not specified 
in their derivation of the hydrodynamics, 
 explicit expressions of the transport coefficients were not given. 
In addition, the Onsager's reciprocal theorem was imposed a priori to determine the ansatz for the distribution function.
In contrast,  the microscopic forms of all the transport coefficients 
are given explicitly in terms of the collision operator in the present work.
Furthermore, it is found that the very microscopic expressions of the transport coefficients have so nice properties that
they readily lead to not only the positivity of the entropy production rate 
but also the Onsager's reciprocal theorem, which we shall prove in the next section.

%---------------------------------------------------------------------------%

%%%%%%%%%%%%%%%%%%%%%%%%%%%%%%%%%%%%%%%%%%%%%%%%%%%%%%%%%%%%%%%%%%%%%

%%%%%%%%%%%%%%%%%%%%%%%%%%%%%%%%%%%%%%%%%%%%%%%%%%%%%%%%%%%%%%%%%%%%%
%\setcounter{equation}{0}
\section{
  Discussions
}
\label{sec:sec5}

We now examine the basic properties of the resultant hydrodynamic equations
(\ref{eq:balanceeqbyRG}) and (\ref{eq:relax1})-(\ref{eq:relax3}) focusing on the entropy production rate.
Since the one-body distribution function \eqref{eq:envelope} is an approximate solution of the Boltzmann equation valid up to $\epsilon^2$ in a global domain, the positivity of the entropy production rate and the Onsager's reciprocal theorem must be satisfied as the Boltzmann theory \cite{de1980relatlvlatlc}.
In this section, we check that these properties are indeed satisfied by Eq.~\eqref{eq:envelope} up to $\epsilon$, since the terms of $O(\epsilon^2)$ do not contribute to the hydrodynamic equation due to the orthogonality relation \eqref{eq:matching1} and \eqref{eq:matching2}.

%---------------------------------------------------------------------------%
\subsection{
  Positivity of entropy production rate
}

The one-body distribution function \eqref{eq:envelope} up to first-order of $\epsilon$ is converted to the following form:
\begin{align} 
 &f_{k,p_k}(\tau_0=\tau)
 \nonumber\\
 &=f_{k,p_k}^{\mathrm{eq}}(1 + \epsilon \bar{f}_{k,p_k}^{\mathrm{eq}}\Psi_{k,p_k})+\mathcal{O}(\epsilon^2)
 \nonumber\\
 &=\Bigg[\exp\Bigg\{\frac{p_k\cdot u}{T}-\sum_{A=1}^Mq_k^A\frac{\mu_A}{T}
 -\epsilon\Psi_{k,p_k}\Bigg\}-a_k\Bigg]^{-1}+\mathcal{O}(\epsilon^2),
 \label{entropy_prod3}
\end{align}
where $\Psi_{k,p_k}$ is given by Eq.~\eqref{eq:Psi_in_RG}.

Inserting Eq.~\eqref{entropy_prod3} into Eq.~\eqref{entropy_prod1}, we have the entropy production rate in the following form,
\begin{align}
  &\partial_\mu s^\mu 
  \nonumber\\
  &=\epsilon\sum_{k=1}^N\int\mathrm{d}p_k 
  f_{k,p_k}^{\mathrm{eq}}\bar{f}_{k,p_k}^{\mathrm{eq}}\Psi_{k,p_k}
  \Bigg(p_k^\mu p_k^\nu \partial_\mu \frac{u_\nu}{T} 
  \nonumber\\
  &-p_k^\mu \partial_\mu \sum_{A=1}^M q_k^A\frac{\mu_A}{T} 
  -\epsilon p_k^\mu\partial_\mu\Psi_{k,p_k}\Bigg) 
  +\mathcal{O}(\epsilon^3)
  \nonumber\\
  &=\epsilon^2  \delta T^{\mu\nu} \nabla_\mu \frac{u_\nu}{T} 
  -\epsilon^2 \sum_{A=1}^M \delta N_A^{\mu} \nabla_\mu \frac{\mu_A}{T} 
  -\epsilon^2\left<\Psi,\frac{\partial}{\partial\tau}\Psi\right>
  \nonumber\\
  &+\mathcal{O}(\epsilon^3)
  \nonumber\\
  &=\epsilon^2  \Bigg(\frac{1}{T\zeta}\Pi^2 +\frac{1}{2T\eta}\pi^{\mu\nu}\pi_{\mu\nu}
  \nonumber\\
  &- \frac{h^2}{T^2}\sum_{A,B=1}^M (\lambda^{-1})_{AB}J_A^{\mu}J_{B,\mu} \Bigg)
  +\mathcal{O}(\epsilon^3),
  \label{entropy_prod2}
\end{align}
where the relaxation equations in the leading order is given by 
\begin{align}
  \label{eq:constitutive_Pi}
  \Pi
  &= -\zeta\theta- \tau_\Pi \frac{\partial}{\partial\tau}\Pi
  +\mathcal{O}(\epsilon), 
  \\
  \label{eq:constitutive_J}
  J_A^\mu
  &= \sum_{B=1}^{M}\lambda_{AB}\frac{T^2}{h^2} \nabla^\mu \frac{\mu_{B}}{T}
  - \sum_{B=1}^{M}\tau_J^{AB} \Delta^{\mu\rho}\frac{\partial}{\partial\tau}J_{B,\rho}
  \nonumber\\
  &+\mathcal{O}(\epsilon),
  \\
  \label{eq:constitutive_pi}
  \pi^{\mu\nu}
  &= 2\eta\sigma^{\mu\nu}
  - \tau_\pi \Delta^{\mu\nu\rho\sigma}\frac{\partial}{\partial\tau}\pi_{\rho\sigma}
  +\mathcal{O}(\epsilon).
\end{align}
In the second equality of Eq.~\eqref{entropy_prod2}, we have used the following transformation
\begin{align}
 \delta T^{\mu\nu} &= \sum_{k=1}^N\int\mathrm{d}p p_k^\mu p_k^\nu 
 f_{k,p_k}^{\mathrm{eq}}\bar{f}_{k,p_k}^{\mathrm{eq}}\Psi_{k,p_k}
 = -\Delta^{\mu\nu}\Pi+\pi^{\mu\nu},
 \\
 \delta N^{\mu}_A &= \sum_{k=1}^N q_k^A \int\mathrm{d}p\, p_k^\mu 
 f_{k,p_k}^{\mathrm{eq}}\bar{f}_{k,p_k}^{\mathrm{eq}}\Psi_{k,p_k}
 = J_A^{\mu},
\end{align}
and
\begin{align}
 &\left<\Psi,\frac{\partial}{\partial\tau}\Psi\right> 
 \nonumber\\
 &=\Pi\frac{1}{T\zeta}\frac{\partial}{\partial\tau}\Pi
 +\frac{h^2}{T^2}\sum_{A,B,C}J_A^\mu (\lambda^{-1})_{AB}\tau_J^{BC}\frac{\partial}{\partial\tau}J_{C,\mu}
 \nonumber\\
 &+\pi^{\mu\nu}\frac{1}{2T\eta}\frac{\partial}{\partial\tau}\pi_{\mu\nu}
 +O({\epsilon}).
\end{align}
In addition, the decomposition of the derivative, $\partial_{\mu}=u_{\mu}\partial/\partial\tau +\epsilon\nabla_{\mu}$, and the orthogonality between the fluid velocity $u_\mu$ and the dissipative contributions to the energy momentum tensor and the particle current, $\delta T^{\mu\nu}$ and $\delta N_A^{\mu}$, have been used.

The viscosities $\zeta$ and $\eta$ are positive quantities as remarked in the last section.
Positive definiteness of the third term of Eq.~\eqref{entropy_prod2} is proved as follows:
\begin{align}
 &- \frac{h^2}{T^2}\sum_{A,B=1}^M (\lambda^{-1})_{AB}J_A^{\mu}J_{B,\mu}
 \nonumber\\
 &=- \frac{h^2}{T^2}\sum_{A,B=1}^M \lambda_{AB}(\lambda^{-1}J^{\mu})_A(\lambda^{-1}J_{\mu})_B
 \nonumber\\
 &=-\frac{h^2}{T^2}\sum_{A,B=1}^M \langle\hat{J}_A^{\nu},L^{-1}\hat{J}_{B,\nu}\rangle 
 (\lambda^{-1}J^{\mu})_A(\lambda^{-1}J_{\mu})_B
 \nonumber\\
 &=\frac{h^2}{T^2} \left< \sum_{A=1}^M U\hat{J}_A^{\nu}(\lambda^{-1}J^{\mu})_A,
 \ \sum_{B=1}^M U\hat{J}_{B,\nu}(\lambda^{-1}J_{\mu})_B\right>
 \nonumber\\
 &>0,
\end{align}
where we have substituted \eqref{eq:TC2byRG} in the first equality and used the Cholesky decomposition in the second equality.
As a result, Eq.~\eqref{entropy_prod2} explicitly shows the positivity of the entropy production rate. 

%---------------------------------------------------------------------------%

%---------------------------------------------------------------------------%
\subsection{
  Onsager's reciprocal theorem
}

Next, we consider the Onsager's reciprocal theorem, which is originated from the time-reversal symmetry of the underlying microscopic theory \cite{onsager1931reciprocal1,onsager1931reciprocal2}. 

The statement of the theorem is as follows.
Near equilibrium, the entropy-production rate takes the following form:
\begin{align}
 \partial_{\mu}s^{\mu}=\sum_i J_iX_i.
\end{align}
where $J_i$ is dissipative currents.
Generally, the sum is taken over any tensor structure of $J_i$ and $X_i$.
In the linear approximation, $J_i$ is given by the following linear combination:
\begin{align}
 J_i=\sum_j \gamma_{ij}X_j,
\end{align}
where all $X_j$ in the sum have the same tensor structure as $J_i$ and the coefficients $\gamma_{ij}$ are scalar quantities.
The Onsager's theorem asserts the symmetry property of the coefficients:
\begin{align}
 \label{eq:Onsager}
 \gamma_{ij}=\gamma_{ji}.
\end{align}

From the expression of the entropy production rate \eqref{entropy_prod2}-\eqref{eq:constitutive_pi}, one readily confirms that the relation \eqref{eq:Onsager} is satisfied in the present case by identifying $J_i=J_A^{\mu}$, $X_i=-(1/h)\nabla^{\mu}(\mu_A/T)$, and $\gamma_{ij}=-\lambda_{AB}T^2$, because $\lambda_{AB}=\lambda_{BA}$ from the microscopic expression of the transport coefficients \eqref{eq:TC2byRG}.
Since this relations are originated from the time-reversal symmetry of the underlying microscopic theory, this confirmation supports the validity of our treatment of the multi-component fluid: Note that the collision integral is given in terms of the microscopic transition rates that respect the time-reversal symmetry.

%---------------------------------------------------------------------------%

%%%%%%%%%%%%%%%%%%%%%%%%%%%%%%%%%%%%%%%%%%%%%%%%%%%%%%%%%%%%%%%%%%%%%

%%%%%%%%%%%%%%%%%%%%%%%%%%%%%%%%%%%%%%%%%%%%%%%%%%%%%%%%%%%%%%%%%%%%%
%\setcounter{equation}{0}
\section{
  Summary and concluding remarks
}
\label{sec:sec6}

In this paper,
we have derived
the second-order hydrodynamic equation for the reactive multi-component systems
 from the relativistic Boltzmann equation based on the RG method. 
%\sout{without making the relaxation-time approximation}.

We have treated the system composed 
of $N$ species with $M$ conserved currents. 
In this case, the hydrodynamic variables 
are composed of $M+4$ moments -- local temperature, fluid velocity, 
and local chemical potentials conjugate to each conserved charges,
 which correspond to the conservation of energy, momentum,
 and M conserved currents. Similarly, the excited $\mathrm{P_1}$-space 
that is incorporated in the second-order hydrodynamics is spanned
 by $3M+6$ vectors because there are dissipative currents due 
to the difference of the diffusion velocity of $M$ currents. 
The characteristic features of the multi-component fluid
 are the cross-correlation effects. 
For instance, $\lambda_{AB}$ in Eq.~\eqref{eq:TC2byRG}, $\tau^{AB}_J$ 
in Eq.~\eqref{eq:RT2byRG}, and so on arose from the correlation between
 microscopic currents associated with the different conserved charges. 
Such coefficients generate the correlation between the diffusions associated
 with different conserved charges.
It is also noteworthy that we have checked the positivity of the entropy production
 rate and  the Onsager's reciprocal theorem are naturally satisfied thanks to our 
adequate derivation of the multi-component hydrodynamics. Since we have faithfully 
solve the multi-component Boltzmann equation, 
these confirmation are also guaranteed by the Boltzmann theory.
Furthermore,
it is worth emphasizing that the form of the resultant relativistic hydrodynamic equation 
is of the same form as that derived in Ref.~\cite{Monnai:2010qp} in which 
the positivity of the enetropy production rate
and the reciprocal relations are imposed a priori in the derivation.
It strongly suggests that the 
form of the reactive multi-component relativistic equation has been now established,
provided that the equation should satisfy the physically natural conditions,
i.e., the positivity of the enetropy production rate
and the reciprocal relations, irrespectively
whether microscopic forms of the transport coefficients are given or not.
 
Finally,
we remark that
the following studies are left as future works:
We will take into account the particle creation and annihilation processes to analyze the realistic QGP hydrodynamic model. Such extension will make it possible to predict the observables 
%such as the photon $v_n$ 
quantitatively.

%%%%%%%%%%%%%%%%%%%%%%%%%%%%%%%%%%%%%%%%%%%%%%%%%%%%%%%%%%%%%%%%%%%%%

%%%%%%%%%%%%%%%%%%%%%%%%%%%%%%%%%%%%%%%%%%%%%%%%%%%%%%%%%%%%%%%%%%%%%
\begin{acknowledgements}
We would like to thank A. Monnai, T. Hirano, and T. M. Doi for useful comments.
We also thank K. Ohnishi for his contribution in the early stage of this work.
This work was supported in part 
by the Core Stage Back UP program in Kyoto University,
by the Grants-in-Aid for Scientific Research from JSPS
%from the Japan Society for the Promotion of Science (JSPS)
% (Nos. 20540265, % TK
 (Nos.24340054, % A.Nakamura (Kiban B)
 24540271%, %, % AO (Kiban C)
)  
and by the Yukawa International Program for Quark-Hadron Sciences.
\end{acknowledgements}
%%%%%%%%%%%%%%%%%%%%%%%%%%%%%%%%%%%%%%%%%%%%%%%%%%%%%%%%%%%%%%%%%%%%%

\appendix

%%%%%%%%%%%%%%%%%%%%%%%%%%%%%%%%%%%%%%%%%%%%%%%%%%%%%%%%%%%%%%%%%%%%%
\setcounter{equation}{0}
\section{
  Detailed derivation of explicit form of excited modes
}
\label{sec:app1}

In this section,
we derive the expression of Eq.~\eqref{eq:Psi_in_RG}.
The initial value of the first-order equation belongs to the space which is spanned by $L^{-1}Q_0F^{(0)}$ as discussed in Ref.~\cite{Tsumura:2015fxa}. Then, the calculation of $L^{-1}Q_0F^{(0)}$ can be reduced to that of
\begin{align}
 \label{1st-PF}
 \big[Q_0F^{(0)}\big]_{k,p_k}
 &= \big[F^{(0)}-P_0F^{(0)}\big]_{k,p_k}
 \nonumber\\
 &= F^{(0)}_{k,p_k}- \varphi_{k,p_k}^{\alpha}\eta^{-1}_{\alpha\beta}
 \langle\varphi^{\beta},F^{(0)}\rangle,
\end{align}
with
\begin{align}
 F^{(0)}_{k,p_k}
 = \frac{1}{p_k\cdot u}\Bigg[p_k^{\mu}p_k^{\nu}\nabla_{\mu}\frac{u_{\nu}}{T}
 -p_k^{\mu}\sum_{A=1}^M q_k^A\nabla_{\mu}\frac{\mu_A}{T}\Bigg].
\end{align}
Here, we have used Eq.~\eqref{eq:F0}
and the equilibrium one-body distribution function $f^{\mathrm{eq}}_{k,p_k}=[\mathrm{e}^{(p_k\cdot u-\sum_A q_k^A \mu_A)/T}-a_k]^{-1}$.

To this end,
we introduce the following quantities:
\begin{align}
 \label{eq:Moment}
 M_\ell^k \equiv \int\mathrm{d}p_k f^{\mathrm{eq}}_{k,p_k}\bar{f}^{\mathrm{eq}}_{k,p_k}(p_k\cdot u)^\ell,\,\,\,\ell=0,\,1,\,\cdots.
\end{align}
Using these quantities,
we have the metric $\eta^{\alpha\beta} = \langle \varphi^\alpha,\varphi^\beta\rangle$ as
\begin{align}
 \eta^{\mu\nu}
 &= \sum_{k=1}^N \Big[a_3^k u^{\mu}u^{\nu} + \frac{m_k^2M_1^k-M_3^k}{3} \Delta^{\mu\nu}\Big],\\
 \eta^{\mu\ A+3} &= \eta^{A+3\ \mu} = \sum_{k=1}^N q_k^A M_2^k u^{\mu},\\
 \eta^{A+3\ B+3} &= \sum_{k=1}^N q_k^A q_k^B M_1^k.
\end{align}
Then,
we the inverse metric $\eta_{\alpha\beta}^{-1}$ is written as
\begin{align}
 \label{1st-invmet8}
 \eta^{-1}_{\mu\nu} &= \mathcal{A} u^{\mu}u^{\nu} + \mathcal{B} \Delta^{\mu\nu},
 \\
 \label{1st-invmet9}
 \eta^{-1}_{\mu\ A+3} &= \eta^{-1}_{A+3\ \mu} = \mathcal{C}_A u^{\mu},
 \\
 \label{1st-invmet10}
 \eta^{-1}_{A+3\ B+3} &= \mathcal{D}_{AB},
\end{align}
where each coefficient satisfies
\begin{align}
 &\mathcal{B}=\left(\sum_{k=1}^N \frac{m^2M_1^k-M_3^k}{3}\right)^{-1},
 \\
 &\sum_{k=1}^N M_3^k \mathcal{A}-1+\sum_{A=1}^{M}\sum_{k=1}^N
 q_k^A M_2^k \mathcal{C}_A=0,
 \\
 &\sum_{k=1}^Nq_k^A M_2^k \mathcal{A}+\sum_{B=1}^{M}\sum_{k=1}^N
 q_k^A q_k^B M_1^k \mathcal{C}_B=0,
 \\
 &\sum_{k=1}^Nq_k^A M_2^k\mathcal{C}_B+\sum_{C=1}^{M}\sum_{k=1}^N
 q_k^A q_k^C M_1^k\mathcal{D}_{CB}=\delta_{AB}.
 \end{align}
The inner products $\langle\varphi^{\beta},F^{(0)}\rangle$
are evaluated as follows: 
\begin{align}
 \label{1st-inner2}
 &\langle\varphi^{\mu},F^{(0)}\rangle
 \nonumber\\
 &=\sum_{k=1}^N\Bigg[\frac{m_k^2M_1^k-M_3^k}{3}\Big( -\frac{1}{T^2}\nabla^{\mu}T
 +u^{\mu}\frac{1}{T}\nabla\cdot u \Big)
 \nonumber\\
 &-\frac{m^2M_0^k-M_2^k}{3}\sum_{A=1}^M q_k^A \nabla^{\mu}\frac{\mu_A}{T}\Bigg],\\
 \label{1st-inner3}
 &\langle\varphi^{A+3},F^{(0)}\rangle
 =\sum_{k=1}^N q_k^A \frac{m_k^2M_0^k-M_2^k}{3}\frac{1}{T}\nabla\cdot u,
\end{align}

Inserting the inverse metric \eqref{1st-invmet8}-\eqref{1st-invmet10} and
the inner products \eqref{1st-inner2} and \eqref{1st-inner3} into Eq. \eqref{1st-PF},
we have
\begin{align}
 \label{eq:Q0F0}
 &\big[ Q_0 F^{(0)} \big]_p
 \nonumber\\
 &=\hat{\Pi}_{k,p_k}\left(-\frac{\theta}{T}\right)
 +\sum_{A=1}^{M}\hat{J}_{A,k,p_k}^{\mu}\left(-\frac{1}{h}\nabla_{\mu}\frac{\mu_A}{T}\right)
 +\hat{\pi}_{k,p_k}^{\mu\nu}\frac{\sigma^{\mu\nu}}{T},
\end{align}
with $(\hat{\Pi},\hat{J}^\mu,\hat{\pi}^{\mu\nu})_{k,p_k}\equiv (\Pi,J^\mu,\pi^{\mu\nu})_{k,p_k}/(p_k\cdot u)$ and $\Pi_{k,p_k}$, $J^\mu_{A,k,p_k}$, and $\pi^{\mu\nu}_{k,p_k}$ are given by
\begin{align}
 &\Pi_{k,p_k} 
 \nonumber\\
 &\equiv (p_k\cdot u)^2\Bigg(\mathcal{A}\sum_{l=1}^N\frac{m_l^2M_1^l-M_3^l}{3}
 \nonumber\\
 &+\sum_{l=1}^N\sum_{A=1}^{M}\mathcal{C}_A q_l^A\frac{m_l^2M_0^l-M_2^l}{3}
 +\frac{1}{3} \Bigg)
 \nonumber \\
 &+(p_k\cdot u)\Bigg(\sum_{l=1}^N\mathcal{C}_A q_k^A\frac{m_l^2M_1^l-M_3^l}{3}
 \nonumber\\
 &+\sum_{l=1}^N\sum_{A,B=1}^{M}\mathcal{D}_{AB}q_k^A q_l^B\frac{m_l^2M_0^l-M_2^l}{3}\Bigg)
 -\frac{1}{3}m_k^2,
 \\
 &J_{A,k,p_k}^{\mu} \equiv 
 \left(q_k^A h-\frac{n_A}{n}(p_k\cdot u)\right)\Delta^{\mu\nu}p_{k\nu}  
 \\
 &\pi_{k,p_k}^{\mu\nu} 
 \equiv \Delta^{\mu\nu\rho\sigma}p_{k\rho}p_{k\sigma}.
\end{align}

Finally, we obtain $\Psi$ as a linear combination of these bases as
\begin{align}
  \label{eq:Psi_in_RG}
  \Psi_{k,p_k} &=
  \Bigg[
    \frac{\big[ L^{-1}\hat{\Pi} \big]_{k,p_k}}{\langle \hat{\Pi},L^{-1}\hat{\Pi} {\rangle}}
  \Bigg] \Pi
  \nonumber\\
  &+\Bigg[
    \sum_{A,B=1}^M 3h\big[ L^{-1} \hat{J}_A^\mu \big]_{k,p_k}\langle \hat{J}^\nu, L^{-1}\hat{J}_{\nu} {\rangle}^{-1}_{AB}  
  \Bigg]  J_{B,\mu}
  \nonumber\\
  &+\Bigg[
    \frac{5\big[ L^{-1}\hat{\pi}^{\mu\nu} \big]_{k,p_k}}{\langle \hat{\pi}^{\rho\sigma}, L^{-1}\hat{\pi}_{\rho\sigma}{\rangle}}
  \Bigg] \pi_{\mu\nu}.
\end{align}
Here
we have introduced the following nine vectors as mere coefficients of the basis vectors: 
\begin{align}
  \Pi(\sigma ;\tau_0),\,\,\,J_A^\mu(\sigma ;\tau_0),\,\,\,\pi^{\mu\nu}(\sigma ;\tau_0),
\end{align}
which are identified as the bulk pressure, the heat flow, and the stress tensor, respectively.

%%%%%%%%%%%%%%%%%%%%%%%%%%%%%%%%%%%%%%%%%%%%%%%%%%%%%%%%%%%%%%%%%%%%%

%%%%%%%%%%%%%%%%%%%%%%%%%%%%%%%%%%%%%%%%%%%%%%%%%%%%%%%%%%%%%%%%%%%%%
\setcounter{equation}{0}
\section{
  Solving the Van del Pol equation by RG method
}
\label{sec:app2}

To show the prescription given in Sec.~\ref{sec:sec3}, we apply the RG method to solve the Van del Pol equation:
\begin{align}
 \ddot{x}+x=\epsilon(1-x^2)\dot{x},
 \label{van_eq}
\end{align}
which is the oscillator with a limit cycle, which is a periodic behavior of the solution in the asymptotic regime. By rewriting it in the form of simultaneous first-order differential equation, we have
\begin{align}
 \left(\frac{\mathrm{d}}{\mathrm{d}t}-A\right)\boldsymbol{u}
 =\epsilon\left(
 \begin{array}{c}
  0 \\
  (1-x^2)y
 \end{array}
 \right),
\end{align}
where $y\equiv \dot{x}$ and the following expressions are introduced
\begin{align}
 \boldsymbol{u} \equiv\left(
 \begin{array}{c}
  x \\
  y
 \end{array}
 \right),
 \ \ \ 
 A \equiv
 \begin{pmatrix}
  0 & 1 \\
  -1 & 0
 \end{pmatrix}.
\end{align}
First, we solve this equation perturbatively up to order $\epsilon$. We denote $\boldsymbol{u}(t)$ for the exact solution and $\tilde{\boldsymbol{u}}(t;t_0)$ for the perturbative solution, which is represented as a perturbation series:
\begin{align}
 \tilde{\boldsymbol{u}}(t;t_0)=\tilde{\boldsymbol{u}}_0(t;t_0)+\epsilon\tilde{\boldsymbol{u}}_1(t;t_0)+\mathcal{O}(\epsilon^2),
\end{align}
with $\tilde{\boldsymbol{u}}_i = {}^t(\tilde{x}_i,\tilde{y}_i)$.
We take the initial value as the exact solution which is also expanded as
\begin{align}
 \tilde{\boldsymbol{u}}(t_0;t_0)=\boldsymbol{u}(t_0)=\boldsymbol{u}_0(t_0)+\epsilon \boldsymbol{u}_1(t_0)+\mathcal{O}(\epsilon^2).
\end{align}

The zeroth-order equations with respect to $\epsilon$ is solved to be
\begin{align}
 \tilde{\boldsymbol{u}}_0(t;t_0)
 =C(t_0)\mathrm{e}^{\mathrm{i}t}\boldsymbol{U}_+ + C^*(t_0)\mathrm{e}^{-\mathrm{i}t}\boldsymbol{U}_-,
\end{align}
with the initial value
\begin{align}
 \tilde{\boldsymbol{u}}_0(t=t_0;t_0)
 &=\boldsymbol{u}_0(t_0)
 \nonumber\\
 &=C(t_0)\mathrm{e}^{\mathrm{i}t_0}\boldsymbol{U}_+ + C^*(t_0)\mathrm{e}^{-\mathrm{i}t_0}\boldsymbol{U}_-,
\end{align}
where $C(t_0)$ is a complex integration constant and $C^*(t_0)$ is its complex conjugate. $\boldsymbol{U}_\pm={}^t(1,\pm\mathrm{i})$ are eigenvectors of the matrix $A$ satisfying $A\boldsymbol{U}_\pm=\pm\mathrm{i}\boldsymbol{U}_\pm$.

The first-order equation with respect to $\epsilon$ reads
\begin{align}
 \left(\frac{\mathrm{d}}{\mathrm{d}t}-A\right)\boldsymbol{u}
 =(1-\tilde{x}_0^2)\tilde{y}_0\frac{1}{2\mathrm{i}}(\boldsymbol{U}_+-\boldsymbol{U}_-).
\end{align}
This equation is solved to be
\begin{align}
 \tilde{\boldsymbol{u}}_1(t;t_0)
 =-\frac{1}{2}(f_+(t;t_0)\boldsymbol{U}_+ + f_-(t;t_0)\boldsymbol{U}_-),
\end{align}
with the definitions of the following quantities 
\begin{align}
 f_+(t;t_0)&\equiv
 \frac{C(t_0)^3}{2\mathrm{i}}\mathrm{e}^{3\mathrm{i}t}
 +\frac{C^*(t_0)^3}{4\mathrm{i}}\mathrm{e}^{-3\mathrm{i}t}
 \nonumber\\
 &-\frac{C^*(t_0)-C(t_0)C^*(t_0)^2}{2\mathrm{i}}\mathrm{e}^{-\mathrm{i}t}
 \nonumber\\
 &-(t-t_0)(C(t_0)-C(t_0)^2C^*(t_0))\mathrm{e}^{\mathrm{i}t},
 \\ 
 f_-(t;t_0)&\equiv
 -\frac{C(t_0)^3}{4\mathrm{i}}\mathrm{e}^{3\mathrm{i}t}
 -\frac{C^*(t_0)^3}{2\mathrm{i}}\mathrm{e}^{-3\mathrm{i}t}
 \nonumber\\
 &+\frac{C(t_0)-C(t_0)^2C^*(t_0)}{2\mathrm{i}}\mathrm{e}^{\mathrm{i}t}
 \nonumber\\
 &-(t-t_0)(C^*(t_0)-C(t_0)C^*(t_0)^2)\mathrm{e}^{-\mathrm{i}t}.
\end{align}
An initial value reads:
\begin{align}
 \tilde{\boldsymbol{u}}_1(t=t_0;t_0)
 &=\boldsymbol{u}_1(t_0)
 \nonumber\\
 &=-\frac{1}{2}(f_+(t_0;t_0)\boldsymbol{U}_+ + f_-(t_0;t_0)\boldsymbol{U}_-).
\end{align}
Here, we have added the zeroth-order solution so that the secular terms vanish at $t=t_0$.
One can find the detailed discussion in Ref.~\cite{Kunihiro:1995zt,Kunihiro:1996rs,Ei:1999pk}.

\begin{figure}
 \begin{center}
 \includegraphics[width=7.5cm]{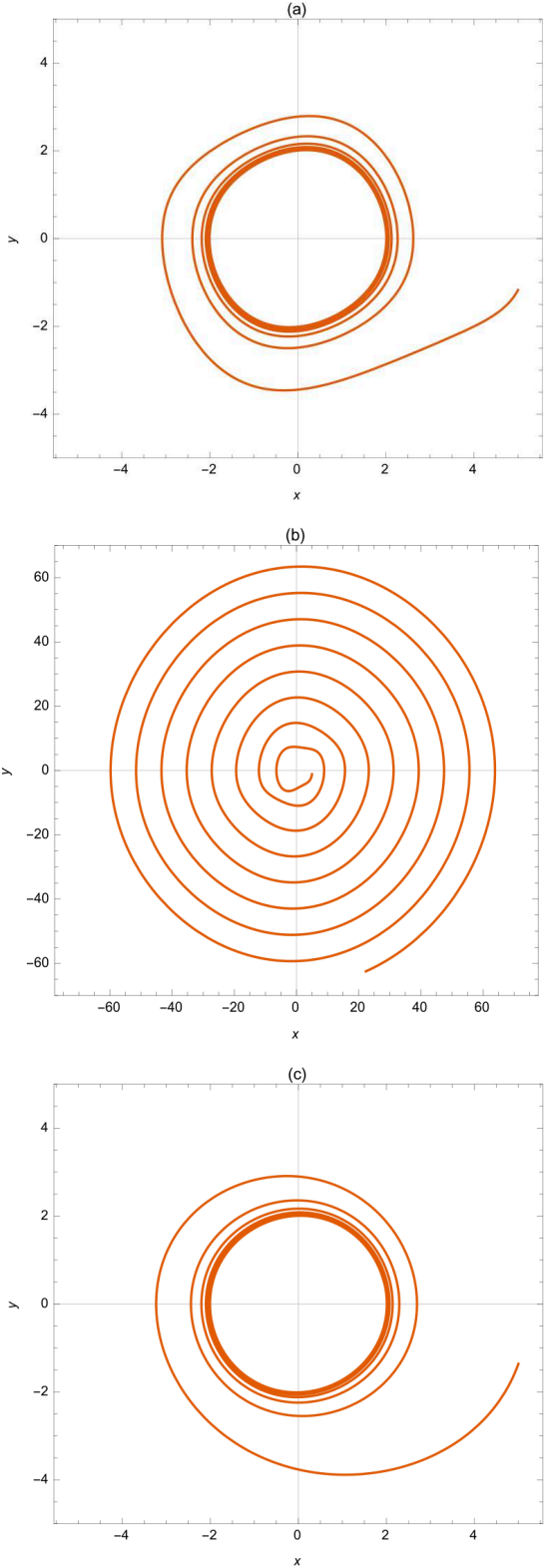}
 \caption{
The orbits
 of the numerical (exact) solution (a), 
the naive perturbative solution (b), 
and the RG-improved solution (c) of the Van del Pol equation \eqref{van_eq}
 in the phase space $(x,y)=(x,\dot{x})$. The initial value is set to $(x,y)=(5.0,-1.4)$
commonly to the three solutions.
}
 \label{van-del-pol}
 \end{center}
\end{figure}

Summing up the solution up to $O(\epsilon)$, we have the perturbative solution and global solution
\begin{align}
 \tilde{\boldsymbol{u}}(t;t_0)
 &=C(t_0)\mathrm{e}^{\mathrm{i}t}\boldsymbol{U}_+ + C^*(t_0)\mathrm{e}^{-\mathrm{i}t}\boldsymbol{U}_-
 \nonumber\\
 &-\epsilon\frac{1}{2}(f_+(t;t_0)\boldsymbol{U}_+ + f_-(t;t_0)\boldsymbol{U}_-),
 \\
 \boldsymbol{u}_G(t)
 &=C(t)\mathrm{e}^{\mathrm{i}t}\boldsymbol{U}_+ + C^*(t)\mathrm{e}^{-\mathrm{i}t}\boldsymbol{U}_-
 \nonumber\\
 &-\epsilon\frac{1}{2}(f_+(t;t)\boldsymbol{U}_+ + f_-(t;t)\boldsymbol{U}_-),
\end{align}
respectively.

Now, one can see that the RG equation:
\begin{align}
 \left.\frac{\mathrm{d}\tilde{\boldsymbol{u}}}{\mathrm{d}t_0}\right|_{t_0=t}=0,
\end{align}
and the Van del Pol equation with the global solution inserted
\begin{align}
 \left(\frac{\mathrm{d}}{\mathrm{d}t}-A\right)\boldsymbol{u}_G
 =\epsilon\left(
 \begin{array}{c}
  0 \\
  (1-x_G^2)y_G
 \end{array}
 \right),
\end{align}
indeed lead to the  same differential equations up to $O(\epsilon)$:
\begin{align}
 \dot{C}(t)&=\frac{\epsilon}{2}\left(C(t)-C(t)^2 C^*(t)\right).
 \label{RG2}
\end{align}
These equations have made the integration constant $C(t_0)$ into the time-dependent variables. It is noteworthy that the right-hand side of the equation are $\mathcal{O}(\epsilon)$, which indicates that we successfully identified the slow variable $C(t)$ and extracted the equations governing the slow dynamics, i.e., describing how the radius of the orbit in the phase space gets relaxed to make a limit cycle. By parametrizing $C(t)$ as $C=(R/2)\exp(\mathrm{i}\theta)$, the amplitude and angular motions may be extracted as
\begin{align}
\label{eq:van-del-pol-R}
 \dot{R}=\epsilon\frac{R}{2}\left(1-\frac{R^2}{4}\right),
 \ \ \ \dot{\theta}=0.
\end{align}
Note that the former of Eq. (\ref{eq:van-del-pol-R}) has a fixed point $R=2$,
i.e., the radius of the ``limit cycle"  toward which the orbit appraoches asymptotically: 
From Fig~\ref{van-del-pol}, we can see that while
the perturbative solution (Fig.~\ref{van-del-pol}(a)) behaves completely different
from the numerical solution (Fig.~\ref{van-del-pol}(b)) due to the secularity,
the RG improved solution (Fig.~\ref{van-del-pol}(c)) 
nicely reproduces the ``limit cycle", i.e., the characteristic property
 of the solution of the Van del Pol equation within the first-order perturbation.

%%%%%%%%%%%%%%%%%%%%%%%%%%%%%%%%%%%%%%%%%%%%%%%%%%%%%%%%%%%%%%%%%%%%%

%%%%%%%%%%%%%%%%%%%%%%%%%%%%%%%%%%%%%%%%%%%%%%%%%%%%%%%%%%%%%%%%%%%%%
\setcounter{equation}{0}
\section{
  Detailed derivation of relaxation equation
}
\label{sec:app3}

In this Appendix,
we present a detailed derivation of the relaxation equation
given by Eqs. (\ref{eq:relax1})-(\ref{eq:relax3}).
The following calculation is based on 
 a lengthy irreducible decomposition of various tensors, 
which are fully utilized in the derivation
 of Eqs.~\eqref{eq:b1}-\eqref{eq:b8} and \eqref{eq:k1}-\eqref{eq:k17}.
An explicit account of the tensor decompositions
 may be seen in a different document.

We introduce the following ``vectors":
\begin{align}
 \hat{\psi}^i_{k,p_k} &\equiv \big\{\hat{\Pi}_{k,p_k},\,\hat{J}^{\mu}_{A,k,p_k},\,\hat{\pi}^{\mu\nu}_{k,p_k}\big\},
 \\
 \psi_i &\equiv \big\{\Pi,\,J_A^{\mu},\,\pi^{\mu\nu}\big\},
 \\
 \hat{\chi}^i &\equiv  \left\{
 \frac{\hat{\Pi}_{k,p_k}}{-T\zeta},\,\frac{\sum_B\hat{J}^{\mu}_{B,k,p_k}(\lambda^{-1})_{BA}}{T^2/h},\,\frac{\hat{\pi}^{\mu\nu}_{k,p_k}}{-2T\eta}\right\},
 \\
  X_i &\equiv  \left\{
 -\zeta\theta,\,\frac{T^2}{h^2}\sum_{B=1}^M \lambda_{AB}\nabla_{\mu}\frac{\mu_B}{T},\,2\eta\sigma^{\mu\nu}\right\},
 \\
 v^{\mu}_{k,p_k} &\equiv \frac{1}{p_k\cdot u}\Delta^{\mu\nu}p_{k,\nu}.
 \label{M-relax3}
\end{align}
Then, Eq. (\ref{eq:ChapB-RHD2}) is converted into the following form:
\begin{align}
  \label{eq:relaxalleq}
  &\langle  \hat{L}^{-1}  \hat{\psi}^{i} ,
  (f^{\mathrm{eq}} \bar{f}^{\mathrm{eq}})^{-1}[\partial_\tau+\epsilon v\cdot\nabla]
  [ f^{\mathrm{eq}}(1 + \epsilon \bar{f}^{\mathrm{eq}} \hat{L}^{-1}  \hat{\psi}^{j}\chi_{j})] \rangle
  \nonumber\\
  &= \epsilon \langle \hat{L}^{-1} \, \hat{\psi}^{i} ,
  \hat{\psi}^{j} \chi_{j} \rangle
  \nonumber\\
  &+ \epsilon^2 \frac{1}{2}\langle \hat{L}^{-1}  \hat{\psi}^{i} ,
  B [ \hat{L}^{-1} \hat{\psi}^{j} \chi_{j}, \hat{L}^{-1} \hat{\psi}^{k} \chi_{k}] \rangle
  + O(\epsilon^3),
\end{align}
Expanding a term in the left-hand side we have
\begin{align}
 &\epsilon \big< L^{-1}\hat{\psi}^i,\hat{\chi}^j\big> X_j
 \nonumber\\
 &+\epsilon\Big<L^{-1}\hat{\psi}^i,(f^{\mathrm{eq}}\bar{f}^{\mathrm{eq}})^{-1}
 \left[\frac{\partial}{\partial\tau}+\epsilon v\cdot\nabla\right]
 f^{\mathrm{eq}}\bar{f}^{\mathrm{eq}}L^{-1}\hat{\chi}^j\Big>\psi_j
 \nonumber\\
 &+\epsilon\big<L^{-1}\hat{\psi}^i,L^{-1}\hat{\chi}^j\big>\frac{\partial}{\partial\tau}\psi_j
 +\epsilon^2\big<L^{-1}\hat{\psi}^i, v^{\alpha}L^{-1}\hat{\psi}^j\big>\nabla_{\alpha}\psi_j
 \nonumber\\
 &=\epsilon\big<L^{-1}\hat{\psi}^i,\hat{\chi}^j\big>\psi_j
 \nonumber\\
 &+\epsilon^2\frac{1}{2}\big<L^{-1}\hat{\psi}^i, B[L^{-1}\hat{\chi}^j, L^{-1}\hat{\chi}^k]\big>\psi_j\psi_k
 + O(\epsilon^3).
 \label{M-relax4}
\end{align}
The coefficients of the first term in the left-hand side and the first and second terms in the right-hand side of Eq.~\eqref{M-relax4} can be written as
\begin{widetext}
\begin{align}
  \langle \hat{L}^{-1} \hat{\psi}^{i},\hat{\chi}^{j} \rangle
  &= \left(
  \begin{array}{ccc}
    \displaystyle{
    \frac{\langle \hat{L}^{-1} \hat{\Pi},\hat{\Pi} \rangle}{-T\zeta}
    }
    &
    \displaystyle{
    \sum_C\frac{\langle \hat{L}^{-1} \hat{\Pi},\hat{J}_C^\rho \rangle}{T^2/h}(\lambda^{-1})_{CB}
    }
    &
    \displaystyle{
    \frac{\langle \hat{L}^{-1} \hat{\Pi},\hat{\pi}^{\rho\sigma} \rangle}{-2T\eta}
    }
    \\
    \displaystyle{
    \frac{\langle \hat{L}^{-1} \hat{J}_A^\mu,\hat{\Pi} \rangle}{-T\zeta}
    }
    &
    \displaystyle{
    \sum_C \frac{\langle \hat{L}^{-1} \hat{J}_A^\mu,\hat{J}_C^\rho \rangle}{T^2/h}(\lambda^{-1})_{CB}
    }
    &
    \displaystyle{
    \frac{\langle \hat{L}^{-1} \hat{J}_A^\mu,\hat{\pi}^{\rho\sigma} \rangle}{-2T\eta}
    }
    \\
    \displaystyle{
    \frac{\langle \hat{L}^{-1} \hat{\pi}^{\mu\nu},\hat{\Pi} \rangle}{-T\zeta}
    }
    &
    \displaystyle{
    \sum_C\frac{\langle \hat{L}^{-1} \hat{\pi}^{\mu\nu},\hat{J}_B^\rho \rangle}{T^2/h}(\lambda^{-1})_{CB}
    }
    &
    \displaystyle{
    \frac{\langle \hat{L}^{-1} \hat{\pi}^{\mu\nu},\hat{\pi}^{\rho\sigma} \rangle}{-2T\eta}
    }
  \end{array}
  \right)
  \nonumber\\
  &= \left(
  \begin{array}{ccc}
    1 & 0 & 0\\
    0 & h\delta_{AB}\Delta^{\mu\rho} & 0\\
    0 & 0 & \Delta^{\mu\nu\rho\sigma}
  \end{array}
  \right),\\
  \langle \hat{L}^{-1} \hat{\psi}^{i},\hat{L}^{-1}\,\hat{\chi}^{j} \rangle
  &= \left(
  \begin{array}{ccc}
    \displaystyle{
    \frac{\langle \hat{L}^{-1} \hat{\Pi},\hat{L}^{-1} \hat{\Pi} \rangle}{-T\zeta}
    }
    &
    \displaystyle{
    \sum_C\frac{\langle \hat{L}^{-1} \hat{\Pi},\hat{L}^{-1} \hat{J}_C^\rho \rangle}{T^2/h}(\lambda^{-1})_{CB}
    }
    &
    \displaystyle{
    \frac{\langle \hat{L}^{-1} \hat{\Pi},\hat{L}^{-1} \hat{\pi}^{\rho\sigma} \rangle}{-2T\eta}
    }
    \\
    \displaystyle{
    \frac{\langle \hat{L}^{-1} \hat{J}_A^\mu,\hat{L}^{-1} \hat{\Pi} \rangle}{-T\zeta}
    }
    &
    \displaystyle{
    \sum_C\frac{\langle \hat{L}^{-1} \hat{J}_A^\mu,\hat{L}^{-1} \hat{J}_C^\rho \rangle}{T^2/h}(\lambda^{-1})_{CB}
    }
    &
    \displaystyle{
    \frac{\langle \hat{L}^{-1} \hat{J}_A^\mu,\hat{L}^{-1} \hat{\pi}^{\rho\sigma} \rangle}{-2T\eta}
    }
    \\
    \displaystyle{
    \frac{\langle \hat{L}^{-1} \hat{\pi}^{\mu\nu},\hat{L}^{-1} \hat{\Pi} \rangle}{-T\zeta}
    }
    &
    \displaystyle{
    \sum_C\frac{\langle \hat{L}^{-1} \hat{\pi}^{\mu\nu},\hat{L}^{-1} \hat{J}_C^\rho \rangle}{T^2/h}(\lambda^{-1})_{CB}
    }
    &
    \displaystyle{
    \frac{langle \hat{L}^{-1} \hat{\pi}^{\mu\nu},\hat{L}^{-1} \hat{\pi}^{\rho\sigma} \rangle}{-2T\eta}
    }
  \end{array}
  \right)\nonumber\\
  &= \left(
  \begin{array}{ccc}
    -\tau_\Pi & 0 & 0\\
    0 & -h\tau_J^{AB}\Delta^{\mu\rho} & 0\\
    0 & 0 & -\tau_\pi \Delta^{\mu\nu\rho\sigma}
  \end{array}
  \right),\\
  \langle \hat{L}^{-1} \hat{\psi}^{i},v^\alpha \hat{L}^{-1}\,\hat{\chi}^{j} \rangle
  &= \left(
  \begin{array}{ccc}
    \displaystyle{
    \frac{\langle \hat{L}^{-1} \hat{\Pi},v^\alpha \hat{L}^{-1} \hat{\Pi} \rangle}{-T\zeta}
    }
    &
    \displaystyle{
    \sum_C\frac{\langle \hat{L}^{-1} \hat{\Pi},v^\alpha \hat{L}^{-1} \hat{J}_C^\rho \rangle}{T^2/h}(\lambda^{-1})_{CB}
    }
    &
    \displaystyle{
    \frac{\langle \hat{L}^{-1} \hat{\Pi},v^\alpha \hat{L}^{-1} \hat{\pi}^{\rho\sigma} \rangle}{-2T\eta}
    }
    \\
    \displaystyle{
    \frac{\langle \hat{L}^{-1} \hat{J}_A^\mu,v^\alpha \hat{L}^{-1} \hat{\Pi} \rangle}{-T\zeta}
    }
    &
    \displaystyle{
    \sum_C\frac{\langle \hat{L}^{-1} \hat{J}_aA^\mu,v^\alpha \hat{L}^{-1} \hat{J}_C^\rho \rangle}{T^2/h}(\lambda^{-1})_{CB}
    }
    &
    \displaystyle{
    \frac{\langle \hat{L}^{-1} \hat{J}_A^\mu,v^\alpha \hat{L}^{-1} \hat{\pi}^{\rho\sigma} \rangle}{-2T\eta}
    }
    \\
    \displaystyle{
    \frac{\langle \hat{L}^{-1} \hat{\pi}^{\mu\nu},v^\alpha \hat{L}^{-1} \hat{\Pi} \rangle}{-T\zeta}
    }
    &
    \displaystyle{
    \sum_C\frac{\langle \hat{L}^{-1} \hat{\pi}^{\mu\nu},v^\alpha \hat{L}^{-1} \hat{J}_C^\rho \rangle}{T^2/h}(\lambda^{-1})_{CB}
    }
    &
    \displaystyle{
    \frac{\langle \hat{L}^{-1} \hat{\pi}^{\mu\nu},v^\alpha \hat{L}^{-1} \hat{\pi}^{\rho\sigma} \rangle}{-2T\eta}
    }
  \end{array}
  \right)\nonumber\\
  &= \left(
  \begin{array}{ccc}
    0 & -h\ell_{\Pi J}^B\Delta^{\alpha\rho} & 0\\
    -\ell_{J \Pi}^A\Delta^{\mu\alpha} & 0 & -\ell_{J \pi}^A\Delta^{\mu\alpha\rho\sigma}\\
    0 & -\ell_{\pi J}^B\Delta^{\mu\nu\rho\alpha} & 0
  \end{array}
  \right),
\end{align}
\end{widetext}
where we have introduced the relaxation times:
\begin{align}
 &\tau_{\Pi}\equiv \frac{1}{T\zeta}\big<L^{-1}\hat{\Pi},L^{-1}\hat{\Pi}\big>,
 \\
 &\tau_{J}^{AB}\equiv -\frac{1}{3T^2}\sum_{C=1}^{M}\big<L^{-1}\hat{J}_A^{\mu},L^{-1}\hat{J}_{C,\mu}\big>(\lambda^{-1})_{CB},
 \\
 &\tau_{\pi}\equiv \frac{1}{10T\eta}\big<L^{-1}\hat{\pi}^{\mu\nu},L^{-1}\hat{\pi}^{\mu\nu}\big>,
\end{align}
and the relaxation length:
\begin{align}
 &\ell_{\Pi J}^A\equiv -\frac{1}{3T^2}\sum_{B=1}^{M}(\lambda^{-1})_{AB}\big<L^{-1}\hat{\Pi},v^{\mu}L^{-1}\hat{J}_{B,\mu}\big>,
 \\
 &\ell_{J\Pi}^A\equiv \frac{1}{T\zeta}\big<L^{-1}\hat{J}_A^{\mu},v_{\mu}L^{-1}\hat{\Pi}\big>,
 \\
 &\ell_{J\pi}^A\equiv \frac{1}{10T\eta}\big<L^{-1}\hat{J}_A^{\mu},v^{\nu}L^{-1}\hat{\pi}_{\mu\nu}\big>,
 \\
 &\ell_{\pi J}^A\equiv -\frac{1}{5T^2}\sum_{B=1}^{M}(\lambda^{-1})_{AB}
 \big<L^{-1}\hat{\pi}^{\mu\nu},v_{\mu}L^{-1}\hat{J}_{B,\nu}\big>.
\end{align}
Then, the last term in the right-hand side of Eq.~\eqref{M-relax4} is written as
\begin{align}
 &-\frac{1}{2}\big<L^{-1}\hat{\Pi},B[L^{-1}\hat{\chi}^j, L^{-1}\hat{\chi}^k]\big>\psi_j\psi_k
 \nonumber\\
 &=b_{\Pi\Pi\Pi}\Pi^2+\sum_{A,B=1}^{M}b_{\Pi JJ}^{AB}J_A^{\rho}J_{B,\rho}+b_{\Pi\pi\pi}\pi^{\rho\sigma}\pi_{\rho\sigma},
 \\[5pt]
 &-\frac{1}{2}\big<L^{-1}\hat{J}_A^{\mu},B[L^{-1}\hat{\chi}^j, L^{-1}\hat{\chi}^k]\big>\psi_j\psi_k
 \nonumber\\
 &=\sum_{B=1}^{M}(b_{J\Pi J}^{AB}\Pi J_B^{\mu}+b_{JJ\pi}^{AB}J_{B,\rho}\pi_{\rho\mu}),
 \\[5pt]
 &-\frac{1}{2}\big<L^{-1}\hat{\pi}^{\mu\nu},B[L^{-1}\hat{\chi}^j, L^{-1}\hat{\chi}^k]\big>\psi_j\psi_k
 \nonumber\\
 &=b_{\pi\Pi\pi}\Pi\pi^{\mu\nu}
 +\sum_{A,B=1}^{M}b_{\pi JJ}^{AB}\Delta^{\mu\nu\rho\sigma}J_{A,\rho}J_{B,\sigma}
 \nonumber\\
 &+b_{\pi\pi\pi}{\pi_{\rho}}^{\lambda}\pi_{\lambda\sigma},
\end{align}

where the transport coefficients are defined by
\begin{align}
 \label{eq:b1}
 b_{\Pi\Pi\Pi} &\equiv -\frac{\big<L^{-1}\hat{\Pi},B[L^{-1}\hat{\Pi}, L^{-1}\hat{\Pi}]\big>}{2(T\zeta)^2},
 \\
 b_{\Pi JJ}^{AB} &\equiv 
 -\sum_{C,D=1}^M \frac{\big<L^{-1}\hat{\Pi},B[L^{-1}\hat{J}_C^{\mu}, L^{-1}\hat{J}_{D,\mu}]\big>}{6(T^2/h)^2}
 \nonumber\\
 &\times(\lambda^{-1})_{CA}(\lambda^{-1})_{DB},
 \\
 b_{\Pi\pi\pi} &\equiv -\frac{\big<L^{-1}\hat{\Pi},B[L^{-1}\hat{\pi}^{\mu\nu}, L^{-1}\hat{\pi}_{\mu\nu}]\big>}{10(2T\eta)^2},
 \\
 b_{J\Pi J}^{AB} &\equiv 
 \sum_{C=1}^M \frac{\big<L^{-1}\hat{J}_A^{\mu},B[L^{-1}\hat{\Pi}, L^{-1}\hat{J}_{C,\mu}]\big>}{3(T\zeta)(T^2/h)}
 (\lambda^{-1})_{CB},
 \\
 b_{JJ\pi} &\equiv 
 \sum_{C,D=1}^M \frac{\big<L^{-1}\hat{J}_A^{\mu},B[L^{-1}\hat{J}_C^{\nu}, L^{-1}\hat{\pi}_{\mu\nu}]\big>}{5(T^2/h)(2T\eta)}
 (\lambda^{-1})_{CB},
 \\
 b_{\pi\Pi\pi} &\equiv -\frac{\big<L^{-1}\hat{\pi}^{\mu\nu},B[L^{-1}\hat{\Pi}, L^{-1}\hat{\pi}_{\mu\nu}]\big>}{5(T\zeta)(T\eta)},
 \\
 b_{\pi JJ} &\equiv 
 -\sum_{C,D=1}^M \frac{\big<L^{-1}\hat{\pi}^{\mu\nu},B[L^{-1}\hat{J}_{C,\mu}, L^{-1}\hat{J}_{D,\nu}]\big>}{10(T^2/h)^2}
 \nonumber\\
 &\times(\lambda^{-1})_{CA}(\lambda^{-1})_{DB},
 \\
 \label{eq:b8}
 b_{\pi\pi\pi} &\equiv -\frac{\big<L^{-1}\hat{\pi}^{\mu\nu},
 B[L^{-1}{\hat{\pi}_{\mu}}^{\lambda},L^{-1}\hat{\pi}_{\lambda\nu}]\big>}{(35/6)(2T\eta)^2}.
\end{align}
Let us rewrite the second term in the right-hand side of Eq.~\eqref{M-relax4}:
\begin{align}
 &\Big<L^{-1}\hat{\psi}^i,(f^{\mathrm{eq}}\bar{f}^{\mathrm{eq}})^{-1}
 \left[\frac{\partial}{\partial\tau}+\epsilon v\cdot\nabla\right]
 f^{\mathrm{eq}}\bar{f}^{\mathrm{eq}}L^{-1}\hat{\chi}^j\Big>\psi_j
 \nonumber\\
 &=\Big<L^{-1}\hat{\psi}^i,(f^{\mathrm{eq}}\bar{f}^{\mathrm{eq}})^{-1}
 \frac{\partial}{\partial T}[f^{\mathrm{eq}}\bar{f}^{\mathrm{eq}}L^{-1}\hat{\chi}^j]\Big>\psi_j
 \frac{\partial}{\partial\tau}T
 \nonumber\\
 &+\sum_{A=1}^{M}\Big<L^{-1}\hat{\psi}^i,(f^{\mathrm{eq}}\bar{f}^{\mathrm{eq}})^{-1}
 \frac{\partial}{\partial\frac{\mu_A}{T}}[f^{\mathrm{eq}}\bar{f}^{\mathrm{eq}}L^{-1}\hat{\chi}^j]\Big>\psi_j
 \frac{\partial}{\partial\tau}\frac{\mu_A}{T}
 \nonumber\\
 &+\Big<L^{-1}\hat{\psi}^i,(f^{\mathrm{eq}}\bar{f}^{\mathrm{eq}})^{-1}
 \frac{\partial}{\partial u^{\beta}}[f^{\mathrm{eq}}\bar{f}^{\mathrm{eq}}L^{-1}\hat{\chi}^j]\Big>\psi_j
 \frac{\partial}{\partial\tau}u^{\beta}
 \nonumber\\
 &+\epsilon\Big<L^{-1}\hat{\psi}^i,(f^{\mathrm{eq}}\bar{f}^{\mathrm{eq}})^{-1}
 v^{\beta}\frac{\partial}{\partial T}[f^{\mathrm{eq}}\bar{f}^{\mathrm{eq}}L^{-1}\hat{\chi}^j]\Big>\psi_j
 \nabla_{\beta}T
 \nonumber\\
 &+\epsilon\sum_{A=1}^{M}\Big<L^{-1}\hat{\psi}^i,(f^{\mathrm{eq}}\bar{f}^{\mathrm{eq}})^{-1}
 v^{\beta}\frac{\partial}{\partial\frac{\mu_A}{T}}[f^{\mathrm{eq}}\bar{f}^{\mathrm{eq}}L^{-1}\hat{\chi}^j]\Big>\psi_j
 \nabla_{\beta}\frac{\mu_A}{T}
 \nonumber\\
 &+\epsilon\Big<L^{-1}\hat{\psi}^i,(f^{\mathrm{eq}}\bar{f}^{\mathrm{eq}})^{-1}
 v^{\beta}\frac{\partial}{\partial u^{\alpha}}[f^{\mathrm{eq}}\bar{f}^{\mathrm{eq}}L^{-1}\hat{\chi}^j]\Big>\psi_j
 \nabla_{\beta}u^{\alpha}.
 \label{M-relax5}
\end{align}
The temporal derivative of $T$, $\mu_A/T$, and $u^{\mu}$ are rewritten by using the 
balance equations up to the first order with respect to $\epsilon$, which correspond to 
the Euler equations:
\begin{align}
 \frac{\partial}{\partial\tau}n_A
 &=-\epsilon n_A\nabla\cdot u+\mathcal{O}(\epsilon^2),
 \\
 \frac{\partial}{\partial\tau}e
 &=-\epsilon nh\nabla\cdot u+\mathcal{O}(\epsilon^2),
 \\
 \frac{\partial}{\partial\tau}u^{\mu}
 &=\epsilon \frac{1}{nh}\nabla^{\mu}P+\mathcal{O}(\epsilon^2).
\end{align}
These equations can be written as
\begin{align}
 &\sum_{k=1}^N q_k^A a_2^k\frac{1}{T^2}\frac{\partial}{\partial\tau}T
 +\sum_{B=1}^M\sum_{k=1}^N q_k^A q_k^B a_1^k\frac{\partial}{\partial\tau}\frac{\mu_B}{T}
 \nonumber\\
 &=-\epsilon n_A\theta
 +\mathcal{O}(\epsilon^2),
 \label{F_Euler1}
 \\
 &\sum_{k=1}^N a_3^k\frac{1}{T^2}\frac{\partial}{\partial\tau}T
 +\sum_{B=1}^M\sum_{k=1}^N q_k^B a_2^k\frac{\partial}{\partial\tau}\frac{\mu_B}{T}
 \nonumber\\
 &=-\epsilon nh\theta
 +\mathcal{O}(\epsilon^2),
 \label{F_Euler2}
 \\
 &\frac{\partial}{\partial\tau}u^{\mu}
 =\epsilon \frac{1}{T}\nabla^{\mu}T
 +\epsilon \frac{T}{h}\sum_{A=1}^{M}\sum_{k=1}^N x_k q_k^A\nabla^{\mu}\frac{\mu_A}{T}
 +\mathcal{O}(\epsilon^2),
 \label{F_Euler3}
\end{align}
where the definitions of $a_1^k$, $a_2^k$, and $a_3^k$ are given by Eq.~\eqref{eq:Moment}.
Let us define a matrix $\mathcal{E}$ by writing Eqs.~\eqref{F_Euler1} and \eqref{F_Euler2} as 
\begin{align}
 &\mathcal{E}_{A0}\frac{\partial}{\partial\tau}T
 +\sum_{B=1}^M\mathcal{E}_{AB}\frac{\partial}{\partial\tau}\frac{\mu_B}{T}
 =\epsilon\theta
 +\mathcal{O}(\epsilon^2),
 \\
 &\mathcal{E}_{00}\frac{\partial}{\partial\tau}T
 +\sum_{B=1}^M\mathcal{E}_{0B}\frac{\partial}{\partial\tau}\frac{\mu_B}{T}
 =\epsilon\theta
 +\mathcal{O}(\epsilon^2).
\end{align}
Therefore, Eqs.~\eqref{F_Euler1}-\eqref{F_Euler1} can be written as
\begin{align}
 \frac{\partial}{\partial\tau}T
 &=\epsilon\mathcal{I}\theta
 +\mathcal{O}(\epsilon^2),
 \\
 \frac{\partial}{\partial\tau}\frac{\mu_A}{T}
 &=\epsilon\mathcal{I}_A\theta
 +\mathcal{O}(\epsilon^2),
 \\
 \frac{\partial}{\partial\tau}u^{\mu}
 &=\epsilon\frac{1}{T}\nabla^{\mu}T
 +\epsilon\frac{n_AT}{nh}\sum_{A=1}^{M}\nabla^{\mu}\frac{\mu_A}{T}
 +\mathcal{O}(\epsilon^2),
\end{align}
where we have defined the coefficients as $\mathcal{I}\equiv\sum_B(\mathcal{E}^{-1})_{0B}$ and $\mathcal{I}_A\equiv\sum_B(\mathcal{E}^{-1})_{AB}$ for notational simplicity.
Then, Eq.~\eqref{M-relax5} takes the following form:
\begin{widetext}
\begin{align}
 &\Big<L^{-1}\hat{\Pi},(f^{\mathrm{eq}}\bar{f}^{\mathrm{eq}})^{-1}
 \left[\frac{\partial}{\partial\tau}+\epsilon v\cdot\nabla\right]
 f^{\mathrm{eq}}\bar{f}^{\mathrm{eq}}L^{-1}\hat{\chi}^j\Big>\psi_j
 \nonumber\\
 &=\epsilon\Big[
 \kappa_{\Pi\Pi}\Pi \theta
 +\sum_{A=1}^M \kappa^{(1)A}_{\Pi J}J_A^\rho\nabla_{\rho}T
 +\sum_{A,B=1}^M\kappa^{(2)AB}_{\Pi J}J_B^\rho\nabla_{\rho}\frac{\mu_A}{T}
 +\kappa_{\Pi\pi}\pi^{\rho\sigma}\sigma_{\rho\sigma}
 \Big],
 \\[5pt]
 &\Big<L^{-1}\hat{J}^{\mu},(f^{\mathrm{eq}}\bar{f}^{\mathrm{eq}})^{-1}
 \left[\frac{\partial}{\partial\tau}+\epsilon v\cdot\nabla\right]
 f^{\mathrm{eq}}\bar{f}^{\mathrm{eq}}L^{-1}\hat{\psi}^j\Big>\chi_j
 \nonumber\\
 &=\epsilon\Big[
 \kappa^{(1)A}_{J\Pi}\Pi\nabla^{\mu}T
 +\sum_{B=1}^M\kappa^{(2)AB}_{J\Pi}\Pi\nabla^{\mu}\frac{\mu_B}{T}
 +\sum_{B=1}^M\kappa^{(1)AB}_{JJ}J_B^\mu \theta
 +\sum_{B=1}^M\kappa^{(2)AB}_{JJ}J_B^\rho{\sigma^\mu}_\rho
 +\sum_{B=1}^M\kappa^{(3)AB}_{JJ}J_B^\rho{\omega^\mu}_\rho
 \nonumber\\
 &+\kappa^{(1)A}_{J\pi}\pi^{\mu\rho}\nabla_{\rho}T
 +\sum_{B=1}^M\kappa^{(2)AB}_{J\pi}\pi^{\mu\rho}\nabla_{\rho}\frac{\mu_B}{T}
 \Big],
 \\[5pt]
 &\Big<L^{-1}\hat{\pi}^{\mu\nu},(f^{\mathrm{eq}}\bar{f}^{\mathrm{eq}})^{-1}
 \left[\frac{\partial}{\partial\tau}+\epsilon v\cdot\nabla\right]
 f^{\mathrm{eq}}\bar{f}^{\mathrm{eq}}L^{-1}\hat{\psi}^j\Big>\chi_j
 \nonumber\\
 &=\epsilon\Big[
 \kappa_{\pi\Pi} \Pi \sigma^{\mu\nu}
 + \sum_{A=1}^M\kappa^{(1)A}_{\pi J}J_A^{\langle\mu}\nabla^{\nu\rangle} T
 + \sum_{A,B=1}^M\kappa^{(2)AB}_{\pi J}J_A^{\langle\mu}\nabla^{\nu\rangle} \frac{\mu_B}{T}
 + \kappa^{(1)}_{\pi\pi}\pi^{\mu\nu} \theta
 + \kappa^{(2)}_{\pi\pi} \pi^{\rho\langle\mu}{\sigma^{\nu\rangle}}_\rho
 + \kappa^{(3)}_{\pi\pi} \pi^{\rho\langle\mu}{\omega^{\nu\rangle}}_\rho
 \Big],
\end{align}
where we have used $\nabla_{\mu}u_{\nu}=\sigma_{\mu\nu}+\Delta_{\mu\nu}\theta/3
+\omega_{\mu\nu}$ with the vorticity $\omega_{\mu\nu}\equiv
(\nabla_{\mu}u_{\nu}-\nabla_{\nu}u_{\mu})/2$, and the transport coefficients are defined 
as follows:
\begin{align}
 \label{eq:k1}
 \kappa_{\Pi\Pi}
 &\equiv 
 \Big<L^{-1}\hat{\Pi},(f^{\mathrm{eq}}\bar{f}^{\mathrm{eq}})^{-1}
 \left[\mathcal{I}\frac{\partial}{\partial T}
 +\sum_{A=1}^M\mathcal{I}_A\frac{\partial}{\partial\frac{\mu_A}{T}}
 +\frac{1}{3}v^\mu \frac{\partial}{\partial u^\mu}\right]
 \frac{f^{\mathrm{eq}}\bar{f}^{\mathrm{eq}}L^{-1}\hat{\Pi}}{-T\zeta}\Big>,
 \\
 \kappa_{\Pi J}^{(1)A}
 &\equiv 
 \frac{\Delta^{\mu\nu}}{3}
 \Big<L^{-1}\hat{\Pi},(f^{\mathrm{eq}}\bar{f}^{\mathrm{eq}})^{-1}
 \left[v_\mu \frac{\partial}{\partial T}+\frac{1}{T}\frac{\partial}{\partial u^\mu}\right]
 \sum_{B=1}^M\frac{f^{\mathrm{eq}}\bar{f}^{\mathrm{eq}}L^{-1}\hat{J}_{B,\nu}}{T^2/h}(\lambda^{-1})_{BA}\Big>,
 \\
 \kappa_{\Pi J}^{(2)AB}
 &\equiv 
 \frac{\Delta^{\mu\nu}}{3}
 \Big<L^{-1}\hat{\Pi},(f^{\mathrm{eq}}\bar{f}^{\mathrm{eq}})^{-1}
 \left[v_\mu \frac{\partial}{\partial\frac{\mu_A}{T}}+\frac{n_AT}{nh}\frac{\partial}{\partial u^\mu}\right]
 \sum_{C=1}^M\frac{f^{\mathrm{eq}}\bar{f}^{\mathrm{eq}}L^{-1}\hat{J}_{C,\nu}}{T^2\lambda/h}(\lambda^{-1})_{CB}\Big>,
 \\
 \kappa_{\Pi\pi}
 &\equiv 
 \frac{\Delta^{\mu\nu\rho\sigma}}{5}\Big<L^{-1}\hat{\Pi},(f^{\mathrm{eq}}\bar{f}^{\mathrm{eq}})^{-1}
 v_\mu \frac{\partial}{\partial u^\nu}\frac{f^{\mathrm{eq}}\bar{f}^{\mathrm{eq}}
 L^{-1}\hat{\pi}_{\rho\sigma}}{-2T\eta}\Big>,
 \\
 \kappa_{J\Pi}^{(1)A}
 &\equiv
 \frac{1}{3}
 \Big<L^{-1}\hat{J}_A^\mu,(f^{\mathrm{eq}}\bar{f}^{\mathrm{eq}})^{-1}
 \left[v_\mu\frac{\partial}{\partial T}+\frac{1}{T}\frac{\partial}{\partial u^\mu}\right]
 \frac{f^{\mathrm{eq}}\bar{f}^{\mathrm{eq}}L^{-1}\hat{\Pi}}{-T\zeta}\Big>,
 \\
 \kappa_{J\Pi}^{(2)AB}
 &\equiv 
 \frac{1}{3}
 \Big<L^{-1}\hat{J}_A^\mu,(f^{\mathrm{eq}}\bar{f}^{\mathrm{eq}})^{-1}
 \left[v_\mu \frac{\partial}{\partial\frac{\mu_B}{T}}+\frac{n_BT}{nh}\frac{\partial}{\partial u^\mu}\right]
 \frac{f^{\mathrm{eq}}\bar{f}^{\mathrm{eq}}L^{-1}\hat{\Pi}}{-T\zeta}\Big>,
 \\
 \kappa_{JJ}^{(1)AB}
 &\equiv 
 \frac{1}{3}
 \Big<L^{-1}\hat{J}_A^\mu,(f^{\mathrm{eq}}\bar{f}^{\mathrm{eq}})^{-1}
 \left[\mathcal{I}\frac{\partial}{\partial T}
 +\sum_{D=1}^M\mathcal{I_D}\frac{\partial}{\partial\frac{\mu_D}{T}}\right]
 \sum_{C=1}^M\frac{f^{\mathrm{eq}}\bar{f}^{\mathrm{eq}}L^{-1}\hat{J}_{C,\mu}}{T^2\lambda/h}(\lambda^{-1})_{CB}\Big>,
 \\
 \kappa_{JJ}^{(2)AB}
 &\equiv 
 \frac{\Delta^{\mu\nu\rho\sigma}}{5}
 \Big<L^{-1}\hat{J}_{A,\mu},(f^{\mathrm{eq}}\bar{f}^{\mathrm{eq}})^{-1}
 v_\rho\frac{\partial}{\partial u^{\sigma}}
 \sum_{C=1}^M\frac{f^{\mathrm{eq}}\bar{f}^{\mathrm{eq}}L^{-1}\hat{J}_{C,\nu}}{T^2\lambda/h}(\lambda^{-1})_{CB}\Big>,
 \\
 \kappa_{JJ}^{(3)AB}
 &\equiv 
 \frac{\Omega^{\mu\nu\rho\sigma}}{3}
 \Big<L^{-1}\hat{J}_{A,\mu},(f^{\mathrm{eq}}\bar{f}^{\mathrm{eq}})^{-1}
 v_\rho\frac{\partial}{\partial u^{\sigma}}
 \sum_{C=1}^M\frac{f^{\mathrm{eq}}\bar{f}^{\mathrm{eq}}L^{-1}\hat{J}_{C,\nu}}{T^2\lambda/h}(\lambda^{-1})_{CB}\Big>,
 \\
 \kappa_{J\pi}^{(1)A}
 &\equiv 
 \frac{\Delta^{\mu\nu\rho\sigma}}{5}
 \Big<L^{-1}\hat{J}_{A,\mu},(f^{\mathrm{eq}}\bar{f}^{\mathrm{eq}})^{-1}
 \left[v_\nu \frac{\partial}{\partial T}+\frac{1}{T}\frac{\partial}{\partial u^\nu}\right]
 \frac{f^{\mathrm{eq}}\bar{f}^{\mathrm{eq}}L^{-1}\hat{\pi}_{\rho\sigma}}{-2T\eta}\Big>,
 \\
 \kappa_{J\pi}^{(2)AB}
 &\equiv 
 \frac{\Delta^{\mu\nu\rho\sigma}}{5}
 \Big<L^{-1}\hat{J}_{A,\mu},(f^{\mathrm{eq}}\bar{f}^{\mathrm{eq}})^{-1}
 \left[v_\nu \frac{\partial}{\partial\frac{\mu_B}{T}}+\frac{n_BT}{nh}\frac{\partial}{\partial u^\nu}\right]
 \frac{f^{\mathrm{eq}}\bar{f}^{\mathrm{eq}}L^{-1}\hat{\pi}_{\rho\sigma}}{-2T\eta}\Big>,
 \\
 \kappa_{\pi\Pi}
 &\equiv 
 \frac{\Delta^{\mu\nu\rho\sigma}}{5}
 \Big<L^{-1}\hat{\pi}_{\mu\nu},(f^{\mathrm{eq}}\bar{f}^{\mathrm{eq}})^{-1}
 v_\rho\frac{\partial}{\partial u^{\sigma}}\frac{f^{\mathrm{eq}}\bar{f}^{\mathrm{eq}}
 L^{-1}\hat{\Pi}}{-T\zeta}\Big>,
 \\
 \kappa_{\pi J}^{(1)A}
 &\equiv
 \frac{\Delta^{\mu\nu\rho\sigma}}{5}
 \Big<L^{-1}\hat{\pi}_{\mu\nu},(f^{\mathrm{eq}}\bar{f}^{\mathrm{eq}})^{-1}
 \left[v_\rho \frac{\partial}{\partial T}+\frac{1}{T}\frac{\partial}{\partial u^\rho}\right]
 \sum_{B=1}^M\frac{f^{\mathrm{eq}}\bar{f}^{\mathrm{eq}}L^{-1}\hat{J}_{B,\sigma}}{T^2\lambda/h}(\lambda^{-1})_{BA}\Big>,
 \\
 \kappa_{\pi J}^{(2)}
 &\equiv 
 \frac{\Delta^{\mu\nu\rho\sigma}}{5}
 \Big<L^{-1}\hat{\pi}_{\mu\nu},(f^{\mathrm{eq}}\bar{f}^{\mathrm{eq}})^{-1}
 \left[v_\rho \frac{\partial}{\partial\frac{\mu}{T}}+\frac{T}{h}\frac{\partial}{\partial u^\rho}\right]
 \frac{f^{\mathrm{eq}}\bar{f}^{\mathrm{eq}}L^{-1}\hat{J}_\sigma}{T^2\lambda/h}\Big>,
 \\
 \kappa_{\pi\pi}^{(1)}
 &\equiv -\frac{\Delta^{\mu\nu\rho\sigma}}{5}
 \Big<L^{-1}\hat{\pi}_{\mu\nu},(f^{\mathrm{eq}}\bar{f}^{\mathrm{eq}})^{-1}
 \left[\mathcal{I}\frac{\partial}{\partial T}
 +\sum_{A=1}^M\mathcal{I}_A\frac{\partial}{\partial\frac{\mu_A}{T}}
 +\frac{1}{3}v^\mu \frac{\partial}{\partial u^\mu}\right]
 \frac{f^{\mathrm{eq}}\bar{f}^{\mathrm{eq}}L^{-1}\hat{\pi}_{\rho\sigma}}{-2T\eta}\Big>,
 \\
 \kappa_{\pi\pi}^{(2)}
 &\equiv 
 \frac{12}{35}\Delta^{\mu\nu\gamma\delta}{\Delta^{\rho\sigma\lambda}}_\gamma{\Delta^{\alpha\beta}}_{\lambda\delta}
 \Big<L^{-1}\hat{\pi}_{\mu\nu},(f^{\mathrm{eq}}\bar{f}^{\mathrm{eq}})^{-1}
 v_\alpha \frac{\partial}{\partial u^\beta}\frac{f^{\mathrm{eq}}\bar{f}^{\mathrm{eq}}
 L^{-1}\hat{\pi}_{\rho\sigma}}{-2T\eta}\Big>,
 \\
 \label{eq:k17}
 \kappa_{\pi\pi}^{(3)}
 &\equiv 
 \frac{4}{15}\Delta^{\mu\nu\gamma\delta}{\Delta^{\rho\sigma\lambda}}_\gamma{\Omega^{\alpha\beta}}_{\lambda\delta}
 \Big<L^{-1}\hat{\pi}_{\mu\nu},(f^{\mathrm{eq}}\bar{f}^{\mathrm{eq}})^{-1}
 v_\alpha\frac{\partial}{\partial u^\beta}\frac{f^{\mathrm{eq}}\bar{f}^{\mathrm{eq}}
 L^{-1}\hat{\pi}_{\rho\sigma}}{-2T\eta}\Big>,
\end{align}
where $\Omega^{\mu\nu\rho\sigma}\equiv (\Delta^{\mu\rho}\Delta^{\nu\sigma}-\Delta^{\mu\sigma}\Delta^{\nu\rho})/2$ is the projection operator onto an antisymmetric tensor of rank two.
Substituting the above equations into Eq.~\eqref{M-relax4},
we arrive at 
the explicit form of the relaxation equations:
\begin{align}
  \label{eq:relax1}
  \epsilon\Pi
  &= -\epsilon\zeta\theta- \epsilon\tau_\Pi \frac{\partial}{\partial\tau}\Pi 
  +\epsilon^2\Bigg(- \sum_{a=1}^{M}\ell^a_{\Pi J}\nabla\cdot J_A
  \nonumber\\
  &+\kappa_{\Pi\Pi} \Pi \theta
  +\sum_{A=1}^{M}\kappa^{(1)A}_{\Pi J}J_{A,\rho}\nabla^\rho T
  +\sum_{A,B=1}^{M}\kappa^{(2)BA}_{\Pi J}J_{A,\rho}\nabla^\rho \frac{\mu_B}{T}
  +\kappa_{\Pi\pi}\pi_{\rho\sigma}\sigma^{\rho\sigma}
   \nonumber\\
  &+ b_{\Pi\Pi\Pi}\Pi^2 + \sum_{A,B=1}^{M}b_{\Pi JJ}^{AB}J_A^\rho J_{B,\rho} + b_{\Pi\pi\pi}\pi^{\rho\sigma}\pi_{\rho\sigma}
  \Bigg),
  \\
  \label{eq:relax2}
  \epsilon J_A^\mu
  &= \epsilon\sum_{B=1}^{M}\lambda_{AB}\frac{T^2}{h^2} \nabla^\mu \frac{\mu_{B}}{T}
  - \epsilon\sum_{B=1}^{M}\tau_J^{AB} \Delta^{\mu\rho}\frac{\partial}{\partial\tau}J_{B,\rho}
  +\epsilon^2\Bigg(- \ell^A_{J\Pi}\nabla^\mu \Pi - \ell^A_{J\pi}\Delta^{\mu\rho} \nabla_\nu {\pi^\nu}_\rho
  \nonumber\\
  &+ \kappa^{(1)A}_{J\Pi}\Pi\nabla^\mu T 
  + \sum_{B=1}^{M}\kappa^{(2)AB}_{J\Pi}\Pi\nabla^\mu \frac{\mu_{B}}{T}
  + \sum_{B=1}^{M}\kappa^{(1)AB}_{JJ}J^\mu_B\theta
  + \sum_{B=1}^{M}\kappa^{(2)AB}_{JJ}J_{B,\rho}\sigma^{\mu\rho}+ \kappa^{(3)AB}_{JJ}J_{B,\rho}\omega^{\mu\rho}
  \nonumber\\
  &+ \kappa^{(1)A}_{J\pi}\pi^{\mu\rho}\nabla_\rho T
  + \sum_{B=1}^{M}\kappa^{(2)AB}_{J\pi}\pi^{\mu\rho}\nabla_\rho \frac{\mu_{B}}{T}
  \nonumber\\
  &+ \sum_{B=1}^{M}b^{AB}_{J\Pi J}\Pi J_B^\mu +  \sum_{B=1}^{M}b^{AB}_{JJ\pi}J_{B,\rho}\pi^{\rho\mu}
  \Bigg),
  \\
  \label{eq:relax3}
  \epsilon\pi^{\mu\nu}
  &= \epsilon2\eta\sigma^{\mu\nu}
  - \epsilon\tau_\pi \Delta^{\mu\nu\rho\sigma}\frac{\partial}{\partial\tau}\pi_{\rho\sigma}
  - \epsilon^2\Bigg(\sum_{a=1}^{M}\ell^a_{\pi J}\nabla^{\langle\mu} J^{\nu\rangle}_a 
  \nonumber\\
  &+\kappa_{\pi\Pi}\Pi\sigma^{\mu\nu}
  + \sum_{A=1}^{M}\kappa^{(1)A}_{\pi J}J^{\langle\mu}_A\nabla^{\nu\rangle} T
  + \sum_{A,B=1}^{M}\kappa^{(2)BA}_{\pi J}J^{\langle\mu}_A\nabla^{\nu\rangle} \frac{\mu_B}{T}
  + \kappa^{(1)}_{\pi\pi}\pi^{\mu\nu}\theta
  + \kappa^{(2)}_{\pi\pi}\pi^{\lambda\langle\mu} {\sigma^{\nu\rangle}}_{\lambda}
  + \kappa^{(3)}_{\pi\pi}\pi^{\lambda\langle\mu} {\omega^{\nu\rangle}}_{\lambda}
  \nonumber\\
  &+ b_{\pi\Pi\pi} \Pi \pi^{\mu\nu}
  + \sum_{A,B=1}^{M}b^{AB}_{\pi JJ} J^{\langle\mu}_A J^{\nu\rangle}_B
  + b_{\pi\pi\pi} \pi^{\lambda\langle\mu} {\pi^{\nu\rangle}}_{\lambda}
  \Bigg).
\end{align}
\end{widetext}
After setting $\epsilon=1$, we find that these equations become
Eqs.~(\ref{eq:relax1})-(\ref{eq:relax3}).

%%%%%%%%%%%%%%%%%%%%%%%%%%%%%%%%%%%%%%%%%%%%%%%%%%%%%%%%%%%%%%%%%%%%%

%%%%%%%%%%%%%%%%%%%%%%%%%%%%%%%%%%%%%%%%%%%%%%%%%%%%%%%%%%%%%%%%%%%%%

%%%%%%%%%%%%%%%%%%%%%%%%%%%%%%%%%%%%%%%%%%%%%%%%%%%%%%%%%%%%%%%%%%%%%

{\bf Note added:}
In the course of finalizing  the present manuscript,
we got aware of the paper by 
A. Jaiswal, B. Friman, and K. Redlich (arXiv:1507.02849), in which the second-order hydrodynamics 
for the relativistic fluid composed of quarks, anti-quarks, and gluons 
is derived from the Boltzmann equation 
within the relaxation-time approximation.
It would be interesting to compare their results with ours
to see 
how the simple relaxation-time approximation is useful and valid.


\begin{thebibliography}{59}
\bibitem[{foo()}]{footnote:vector}
\bibinfo{note}{We call functions of $(k,p_k)$ ``vector" in this article}.
\end{thebibliography}
\end{document}